\documentclass[reqno,12pt,a4paper]{article}
\usepackage{amsfonts}
\usepackage{amsmath}
\usepackage{dpdefs}
\usepackage{lscape}
\usepackage{titlesec}
\usepackage{afterpage}

\titleformat{\section}{\normalfont\Large\bfseries\singlespacing}{\thesection}{1em}{}
\titlespacing*{\section}{0pt}{0ex}{0.6\baselineskip}

\floatstyle{onerule}
\restylefloat{table}
\newtheorem{assumption}{Assumption}
\hypersetup{pdfstartview={FitH}, citecolor=blue }
\input{tcilatex}
\begin{document}

\title{\singlespacing Bias Correction of Persistence Measures in Fractionally Integrated
Models}
\author{Simone D. Grose, Gael M. Martin\thanks{%
Corresponding author: Gael Martin, Department of Econometrics and Business
Statistics, Monash University, Clayton, Victoria 3800, Australia. Tel.:
+61-3-9905-1189; fax: +61-3-9905-5474; email: gael.martin@monash.edu.} and
Donald S. Poskitt \\
{\small \emph{Department of Econometrics \& Business Statistics, Monash
University}}}
\maketitle

\begin{abstract}\medskip
\noindent This paper investigates the accuracy of bootstrap-based bias
correction of persistence measures for long memory fractionally integrated
processes. The bootstrap method is based on the semi-parametric sieve
approach, with the dynamics in the long memory process captured by an
autoregressive approximation. With a view to improving accuracy, the sieve
method is also applied to data pre-filtered by a semi-parametric estimate of
the long memory parameter. Both versions of the bootstrap technique are used
to estimate the finite sample distributions of the sample autocorrelation
coefficients and the impulse response coefficients and, in turn, to
bias-adjust these statistics. The accuracy of the resultant estimators in
the case of the autocorrelation coefficients is also compared with that
yielded by analytical bias adjustment methods when available. The basic
sieve technique is seen to yield a reduction in the bias of both persistence
measures. The pre-filtered sieve produces a substantial further reduction in
the bias of the estimated impulse response function, whilst the extra
improvement yielded by pre-filtering in the case of the sample
autocorrelation function is shown to depend heavily on the accuracy of the
pre-filter.

\medskip


\medskip

\end{abstract}
\newpage

\section{Introduction}

Measuring the degree of persistence, or memory, in an economic or financial
time series is crucial for understanding\ the response of the variable to
shocks, in particular to policy-induced shocks. Traditionally, discussion of
persistence has taken place in the context of models that are either
integrated of order zero ($I(0)$) or of order one ($I(1)$), with the most
commonly applied measures in this context being the impulse response and
autocorrelation functions. The focus of this paper is on measuring
persistence in the class of fractionally integrated ($I(d)$) processes
introduced by \cite{granger:joyeux:1980} and \cite{hosking:1981} -- a key
class of models used to capture long memory, or \textit{strong} dependence,
in a wide range of empirical applications.

Long memory $I(d)$ processes can be characterized by the specification%
\begin{equation}
y(t)=\sum_{j=0}^{\infty }\psi (j)\varepsilon (t-j)=\frac{\kappa (z)}{%
(1-z)^{d}}\,\varepsilon (t),  \label{Wold}
\end{equation}%
where $\varepsilon (t)$, $t\in {\mathcal{Z}}$, is a zero mean white noise
process with variance $\sigma ^{2}$, $z$ denotes the lag operator, and the
`short-memory' component, $\kappa (z)=\sum_{j\geq 0}\kappa (j)z,$ is assumed
to satisfy $\sum_{j\geq 0}|\kappa (j)|<\infty $, the transfer function of a
stable, invertible autoregressive moving average (ARMA) process, for
example. The long-run behaviour of this process depends on the fractional
integration parameter $d$. Specifically, for any $d\neq 0$ the impulse
response coefficients $\psi (z)$ in (\ref{Wold}), as well as the
autocovariances of the process, decline at a hyperbolic rate, rather than
the exponential rate typical of an ARMA process. For the empirically
relevant values of $d>0$ the rate of decline is slow enough to preclude
absolute summability for both measures of persistence, leading to the
characterization of $y(t)$ as a `long-memory' process in this case.

While the literature dealing with inference in the context of autoregressive
fractionally integrated moving average (ARFIMA) models is well-developed%
\footnote{%
See \cite{beran:1994}, \cite{doukhan:oppenheim:taqqu:2003} and \cite%
{robinson:2003} for textbook expositions}, some issues remain to be
addressed, including those pertaining to inference about the two persistence
measures. Most notable here is the well-known downward bias of estimates of
the autocorrelation function (ACF) under long memory \citep[][]{hosking:1996}%
, and the impact on inference of the asymptotic non-Gaussianity of the
sample autocorrelations for $d\geq 0.25$. Regarding the bias issue
specifically, while \cite{hosking:1996} provides an asymptotically valid
representation of the bias of the general $k^{th}$-order sample
autocorrelation, it would require estimates of unknown population parameters
to yield a feasible bias-adjustment method, and the sampling properties of
any resultant bias-adjusted estimator remain unknown. The same point holds
for the higher-order result for the bias of the first-order sample
autocorrelation coefficient derived by \cite{lee:ko:2009}. Similarly, whilst
the general problem of producing accurate point and interval estimates of
the impulse response function (IRF) in time series models has prompted
recent investigation (see \citealp{pesavento:rossi:2007,inoue:kilian:2014};
and \citealp{lutkepohl:bystrova:winker:2014}; for recent examples), the
specific issue of IRF inference in long memory ARFIMA processes -- including
that of bias correction -- has to our knowledge only been tackled in \cite%
{baillie:kapetanios:2013}, and remains an under-developed area.

The primary focus of the current paper is on the use of bootstrap methods to
bias correct both persistence measures in the long memory ARFIMA setting. In
the spirit of recent work in \cite{poskitt:2008}, \cite%
{baillie:kapetanios:2013} and \cite{poskitt:grose:martin:2013}, the
semi-parametric sieve bootstrap is the technique of choice, obviating as it
does the need to specify the unknown short-run dynamics in the ARFIMA model.
The sieve works by `whitening' the data using an autoregressive (AR)
approximation, capturing the dynamics of the process in the fitted
autoregression, the order of which increases at a suitable rate with the
sample size. Results presented by \cite{poskitt:2008}, building on earlier
results in \cite{poskitt:2007}, demonstrate that the sieve method produces
error rates that are superior to those of the block bootstrap of \cite%
{kunsch:1989}. Subsequently, \cite{poskitt:grose:martin:2013} have
strengthened these results considerably, with the higher-order improvement
yielded by the sieve method demonstrated using an Edgeworth expansion for a
broad class of statistics that includes both forms of statistics
investigated here. Furthermore, the authors have shown that the rate of
convergence of a modified version of the sieve, in which a consistent
semi-parametric estimator of $d$\ is used to `pre-filter' the data prior to
the application of the sieve algorithm, is equivalent to that associated
with the application of the sieve method to short memory processes\footnote{%
See \cite{choi:hall:2000}. This rate is, in turn, arbitrarily close to the
bootstrap rate of convergence attained for independent data.}.

In the current paper we exploit the theoretical (and numerical) accuracy of
the sieve-based distribution estimates, and extract from those estimated
distributions an appropriate estimate of the bias in the statistics of
interest. The finite sample properties of the bias-adjusted estimators so
produced are then documented via an extensive simulation exercise.
Consistent with the semi-parametric spirit of the exercise, the impulse
response coefficients are produced as the inversion of an autoregression
fitted to the data, rather than as non-linear functions of the parameters of
some fully specified ARFIMA model. The sample autocorrelation coefficients
are calculated using the standard Pearson formula. For both persistence
functions the pre-filtered sieve is illustrated using the `semi-parametric
Gaussian' estimator of $d$ examined by \cite{robinson:1995b}, here referred
to as the `semi-parametric local Whittle' (SPLW) estimator. This estimator
is shown in \cite{poskitt:grose:martin:2013} to satisfy the necessary
conditions for the higher-order convergence properties of the pre-filtered
sieve to obtain. As a proof-of-concept exercise, we also document results
based on the use of the true (unknown) value of $d$ as the pre-filter.

The paper proceeds as follows. Section 2 briefly outlines the methodology
underlying the sieve bootstrap and its use in estimating the sampling
distribution and finite-sample bias of selected persistence measures. For
conciseness we present the more general pre-filtered methodology in detail,
with this technique nesting the `raw' sieve technique when the pre-filtering
step is omitted. Selected results from \cite{poskitt:grose:martin:2013}
detailing the theoretical convergence rates on which the subsequent
bias-adjustment rests are also included. In Section 3 we outline the
properties of the two persistence measures to be bias-adjusted, whilst in
Section 4 the finite sample performance of the bias-corrected estimators in
a variety of settings is assessed via simulation.\vspace{-12pt}

\section{Long-memory processes, autoregressive approximation, and the
pre-filtered sieve bootstrap\label{sieve}}

We assume that $y(t)$ is a linearly regular, covariance-stationary process
with representation as in \eqref{Wold} where the stochastic disturbance and
the impulse response coefficients satisfy the following conditions:

\begin{assumption}
\label{Ass1} The process $\varepsilon (t)$ is ergodic, and
\begin{equation}
E\big[\varepsilon (t)\mid \mathcal{E}_{t-1}\big]=0\quad \text{and\quad }E%
\big[\varepsilon (t)^{2}\mid \mathcal{E}_{t-1}\big]=\sigma ^{2}\,,
\end{equation}%
where $\mathcal{E}_{t}$ denotes the $\sigma $-algebra of events determined
by $\varepsilon (s)$, $s\leq t$. Furthermore, $E[\varepsilon (t)^{4}]<\infty
$.
\end{assumption}

\begin{assumption}
\label{Ass2} The transfer function in the representation of the process $%
y(t) $, namely $k(z)=\sum_{j\geq 0}\psi (j)z^{j}$, is given by $\psi
(z)=\kappa (z)/(1-z)^{d}$ where $|d|<0.5$ and $\kappa (z)$ satisfies $\kappa
(z)\neq 0$, $|z|\leq 1$, and $\sum_{j\geq 0}j|\kappa (j)|<\infty $.
\end{assumption}

Assumption \ref{Ass1} imposes a classical martingale difference structure on
the stochastic disturbance process; the key property of such a process that
underlies the asymptotic results being that a martingale difference is
uncorrelated with any measurable function of its own past. Assumptions \ref%
{Ass1} and \ref{Ass2}, taken together, incorporate a wide class of linear
processes, including the ARFIMA family of models that are the focus of this
work.

Under the martingale difference structure for $\varepsilon (t)$ imposed by
Assumption \ref{Ass1}, the linear predictor $\bar{y}(t)=\sum_{j=1}^{\infty
}\pi (j)y(t-j)$ is the minimum mean squared error predictor (MMSEP) of $y(t)$%
. The MMSEP of $y(t)$ based only on the {finite past} is then
\begin{equation}
\bar{y}_{h}(t)=\sum_{j=1}^{h}\pi _{h}(j)y(t-j)\equiv -\sum_{j=1}^{h}\phi
_{h}(j)y(t-j),  \label{plinh}
\end{equation}%
where we adopt the minor reparameterization from $\pi _{h}$ to $\phi _{h}$
in order to allow us, on also defining $\phi _{h}(0)=1$, to write the
corresponding prediction error as $\varepsilon _{h}(t)=\sum_{j=0}^{h}\phi
_{h}(j)y(t-j)$. The finite-order autoregressive coefficients $\phi
_{h}(1),\ldots ,\phi _{h}(h)$ can, in turn, be deduced from the Yule-Walker
equations, $\sum_{j=0}^{h}\phi _{h}(j)\gamma (j-k)=\delta _{0}(k)\sigma
_{h}^{2}\,$, $k=0,1,\ldots ,h$, where $\gamma (\tau )=\gamma (-\tau
)=E[y(t)y(t-\tau )]$, $\tau =0,1,\ldots $ is the autocovariance function of
the process $y(t)$, $\delta _{0}(k)$ is Kronecker's delta (i.e., $\delta
_{0}(k)=0\;\forall \;k\neq 0$; $\delta _{0}(0)=1$), and
\begin{equation}
\sigma _{h}^{2}=E\left[ \varepsilon _{h}(t)^{2}\right]  \label{Varh}
\end{equation}%
is the prediction error variance associated with $\bar{y}_{h}(t)$ in (\ref%
{plinh}).

The use of the optimal predictor $\bar{y}_{h}(t)$ determined from the
autoregressive model of finite order $h$ is appropriate only if it is a good
approximation to the `infinite-order'\ predictor $\bar{y}(t)$ for
sufficiently large $h$. \cite{poskitt:2007} addresses this very issue under
regularity conditions that admit non-summable processes, proving the
asymptotic validity, and properties, of finite-order autoregressive models
when $h\rightarrow \infty $ with the sample size $T$ at a suitable rate. In
brief, the order-$h$ prediction error $\varepsilon _{h}(t)$ converges to $%
\varepsilon (t)$ in mean-square, the estimated sample-based covariances
converge to their population counterparts -- albeit at a slower rate than
for a conventionally stationary process -- and the least squares and \YW\
estimators of the coefficients of the approximating autoregression are
asymptotically equivalent and consistent. It thus follows %
\citep[see][]{poskitt:2008}, that the sieve bootstrap, which uses an
estimated autoregressive approximation to capture the dynamics of the
process, is a plausible semi-parametric bootstrap technique for long-memory
processes.

Motivated by the theoretical results in \cite{poskitt:grose:martin:2013},
we, in turn, modify this `raw' sieve approach by applying the sieve after
the data has been pre-filtered via a suitable $\sqrt{N}$-consistent
semi-parametric estimator of $d$, where $N$\ increases with $T$\ such that $%
N/T\rightarrow 0$\ as $T\rightarrow \infty $. Details of the both the raw
and pre-filtered sieve bootstrap, including their relevant orders of
accuracy are, as noted earlier, given in \cite{poskitt:grose:martin:2013}.
For convenience, we describe here the basic steps needed to implement the
pre-filtered sieve bootstrap. A brief summary of the relevant convergence
results from \cite{poskitt:grose:martin:2013} then follows in Section \ref%
{converg}.

\subsection{The pre-filtered sieve algorithm\label{pf_alg}}

Suppose that a value $\widehat{d}$ is available such that $\widehat{d}-d\in
N_{\delta }=\{x:|x|<\delta \}$ where $0<\delta <0.5$. For any $d>-1$ let $%
\alpha _{j}^{(d)}$, $j=0,1,2,\ldots $, denote the coefficients of the
binomial expansion of the fractional difference operator, $%
(1-z)^{d}=\sum_{j=0}^{\infty }\alpha _{j}^{(d)}z^{j}$ $=1+\sum_{j=1}^{\infty
}\left( \frac{\Gamma (j-d)}{\Gamma (-d)\Gamma (j+1)}\right) z^{j}$ $%
=1+\sum_{j=1}^{\infty }\left( \prod_{0<k\leq j}\frac{k-1-d}{k}\right) z^{j}$%
. Setting $w(t)=\sum_{j=0}^{t-1}\alpha _{j}^{(d)}y(t-j)$, $t=1,\ldots ,T\,,$
and using the preliminary estimate $\widehat{d}$, pre-filtered sieve
bootstrap realizations of $y(t)$ are generated as follows:

\begin{description}
\item[Step 1.] Calculate the coefficients of the filter $(1-z)^{\widehat{d}}$
and from the empirical data generate the filtered values $\widehat{w}%
(t)=\sum_{j=0}^{t-1}\alpha _{j}^{(\widehat{d})}y(t-j)$, $t=1,\ldots ,T$.

\item[Step 2.] Fit an AR approximation to $\widehat{w}(t)$ and generate a
sieve bootstrap sample $\widehat{w}^{\ast }(t)$, $t=1,\ldots ,T$, of the
filtered data as follows:

\begin{enumerate}
\item \label{s1} Given the filtered series $\widehat{w}(t)$, $t=1,\ldots ,T$%
, calculate the parameter estimates of the $AR(h)$ approximation, denoted by
$\hat{\phi}_{h}(1),\ldots ,\hat{\phi}_{h}(h)$ and $\hat{\sigma}_{h}^{2}$,
and evaluate the residuals, $\hat{\varepsilon}_{h}(t)=\sum_{j=0}^{h}\hat{\phi%
}_{h}(j)\widehat{w}(t-j)$, $t=1,\ldots ,T\,,$ using $\widehat{w}(1-j)=%
\widehat{w}(T-j+1)$, $j=1,\ldots ,h$, as initial values. From $\hat{%
\varepsilon}_{h}(t)$, $t=1,\ldots ,T$, construct the standardized residuals $%
\tilde{\varepsilon}_{h}(t)=(\hat{\varepsilon}_{h}(t)-\bar{\varepsilon}%
_{h})/s_{\hat{\varepsilon}_{h}}$, where $\bar{\varepsilon}%
_{h}=T^{-1}\sum_{t=1}^{T}\hat{\varepsilon}_{h}(t)$ and $s_{\hat{\varepsilon}%
_{h}}^{2}=T^{-1}\sum_{t=1}^{T}(\hat{\varepsilon}_{h}(t)-\bar{\varepsilon}%
_{h})^{2}$.

\item Let $\varepsilon _{h}^{+}(t)$, $t=1,\ldots ,T$, denote a simple random
sample of \textit{i.i.d.} values drawn from $U_{\tilde{\varepsilon}%
_{h},T}(e)=T^{-1}\sum_{t=1}^{T}\mathbf{1}\{\tilde{\varepsilon}_{h}(t)\leq
e\} $, the probability distribution function that places a probability mass
of $1/T$ at each of $\tilde{\varepsilon}_{h}(t)$, $t=1,\ldots ,T$. Set $%
\varepsilon _{h}^{\ast }(t)=\hat{\sigma}_{h}\varepsilon _{h}^{+}(t)$, $%
t=1,\ldots ,T$.

\item Construct the sieve bootstrap realization $\widehat{w}^{\ast
}(1),\ldots ,\widehat{w}^{\ast }(T)$ where $\widehat{w}^{\ast }(t)$ is
generated from the autoregressive process $\sum_{j=0}^{h}\hat{\phi}_{h}(j)%
\widehat{w}^{\ast }(t-j)=\varepsilon _{h}^{\ast }(t)$, $t=1,\ldots ,T\,,$
initiated at $\widehat{w}^{\ast }(1-j)=\widehat{w}(\tau -j+1)$, $j=1,\ldots
,h$, where $\tau $ has the discrete uniform distribution on the integers $%
h,\ldots ,T$.
\end{enumerate}

\item[Step 3.] Using the coefficients of the (inverse) filter $(1-z)^{-%
\widehat{d}}$, construct, for $y(t)$, the corresponding pre-filtered sieve
bootstrap draw, $\widehat{y}^{\ast }(t)=\sum_{j=0}^{t-1}\alpha _{j}^{(-%
\widehat{d})}\widehat{w}^{\ast }(t-j)$, $t=1,\ldots ,T$, from which the
relevant statistics -- the autocorrelation and impulse response coefficients
in this case -- are computed.
\end{description}

The raw bootstrap is nested in the above algorithm. Specifically, it
involves the omission of Steps 1 and 3 above, and the application of Step 2
to the raw data $y(t)$ rather than the pre-filtered series $\widehat{w}(t)$.

By simulating a large number of such bootstrap samples, the empirical
distribution function of any given statistic is produced, representing,
under suitable conditions, a valid approximation to the unknown true
sampling distribution of the statistic in question. Conditional on this
validity, an estimate of bias can be extracted via the bootstrap
distribution, and a bias-corrected statistic thereby produced. The
properties of this technique applied to the statistics of interest here
follow from the convergence results proved in \cite%
{poskitt:grose:martin:2013} and outlined below.

\subsection{Key convergence results\label{converg}}

We begin by highlighting the fact that the process $(1-z)^{\widehat{d}}y(t)=%
\frac{\kappa (z)}{(1-z)^{d-\widehat{d}}}\,\varepsilon (t)$ has fractional
index $d-\widehat{d}$. By the first Theorem of
\citet[Section
2]{poskitt:grose:martin:2013}, the error in the autoregressive approximation
to $\widehat{w}(t)$ will accordingly be of order $O(h\left( \ln T/T\right)
^{1-2|d-\widehat{d}|})$ or smaller, in contrast to the AR approximation
error associated with the \textit{raw} sieve, which is $O(h\left( \ln
T/T\right) ^{1-2d^{\prime }})$ with $d^{\prime }=\max \{0,d\}.$ Thus
pre-filtering can yield increased accuracy depending on the value of $|d-%
\widehat{d}|$. That this (potential) increase in accuracy is transferred to
the pre-filtered sieve bootstrap realizations $\widehat{y}^{\ast }(t)$ of $%
y(t)$, via the sieve bootstrap draws $\widehat{w}^{\ast }(t)$ of $\widehat{w}%
(t)$, and hence to the pre-filtered sieve bootstrap approximation to the
sampling distribution of any given statistic in a suitable class, rests upon
the first Proposition of \citet[Section 4]{poskitt:grose:martin:2013}, the
proof of which is given in that paper. The ultimate consequence of the use
of suitable pre-filtering is an improved rate of convergence for the
bootstrap-based estimate of the relevant sampling distribution, vis-a-vis
the corresponding estimate based on the raw sieve. We summarize those
convergence results briefly as follows.

Denote the relevant statistic as $\mathbf{s}_{T}=(s_{1T},\ldots
,s_{mT})^{\prime }$, where $s_{iT}=s_{i}(y(1),\ldots ,y(T))$, and each $%
s_{i}(\cdot )$ for $i=1,\ldots ,m$ is a suitably smooth function of the time
series values $y(1),\ldots ,y(T)$ that falls within the broad class of
statistics that satisfy the two assumptions specified in Section 3 of \cite%
{poskitt:grose:martin:2013}, a class that includes the sample
autocorrelation and impulse response functions considered in this paper. Let
$\mathbf{s}_{T}^{\ast }$ be defined as for $\mathbf{s}_{T}$ but with the
observed data replaced by $y^{\ast }(1),\ldots ,y^{\ast }(T)$, a realization
obtained from the sieve bootstrap algorithm, so that $\mathbf{s}_{T}^{\ast
}=(s_{1T}^{\ast },\ldots ,s_{mT}^{\ast })^{\prime }$ where $s_{iT}^{\ast
}=s_{i}(y^{\ast }(1),\ldots ,y^{\ast }(T))$. Further define $\mathbf{V}%
_{T}=T^{-1}E\left[ (\mathbf{s}_{T}-E[\mathbf{s}_{T}])(\mathbf{s}_{T}-E[%
\mathbf{s}_{T}])^{\prime }\right] $ and $\bzeta_{T}=\mathbf{V}_{T}^{-1/2}T^{-%
\half}(\mathbf{s}_{T}-E[\mathbf{s}_{T}])$, where $E$ denotes expectation
taken with respect to the original probability space $(\Omega ,\mathfrak{F}%
,P)\ $, and $\mathbf{V}_{T}^{\ast }=T^{-1}E^{\ast }\left[ (\mathbf{s}%
_{T}^{\ast }-E^{\ast }[\mathbf{s}_{T}^{\ast }])(\mathbf{s}_{T}^{\ast
}-E^{\ast }[\mathbf{s}_{T}^{\ast }])^{\prime }\right] $ and $\bzeta%
_{T}^{\ast }=\mathbf{V}_{T}^{\ast -1/2}T^{-\half}(\mathbf{s}_{T}^{\ast
}-E^{\ast }[\mathbf{s}_{T}^{\ast }])$, where $E^{\ast }$ denotes expectation
taken with respect to the (relevant) bootstrap probability space $(\Omega
^{\ast },\mathfrak{F}^{\ast },P^{\ast }).$ Under the relevant conditions
stated in \cite{poskitt:grose:martin:2013} (and with proofs included
therein) it follows that for the raw sieve method
\begin{equation}
\sup_{\mathbf{z}}|\QTR{md}{P}^{\ast }(\widehat{\bzeta}_{T}^{\ast }\leq
\mathbf{z})-\QTR{md}{P}(\bzeta_{T}\leq \mathbf{z})|=O_{p}(T^{-(1-d^{\prime
})+\beta }),  \label{raw_bs}
\end{equation}%
for all $\beta >0$, where $d^{\prime }=\max \{0,d\}$. For the pre-filtered
method, for all pre-filtering estimates $\widehat{d}$ such that $\widehat{d}%
-d\in N_{\delta _{T}}$ where $\delta _{T}\log T\overset{a.s.}{\rightarrow }0$
as $T\rightarrow \infty $,
\begin{equation}
\sup_{\mathbf{z}}|\QTR{md}{P}^{\ast }(\widehat{\bzeta}_{T}^{\ast }\leq
\mathbf{z})-\QTR{md}{P}(\bzeta_{T}\leq \mathbf{z})|=\exp (\delta _{T}\log
T)O_{p}(T^{-1+\beta }),  \label{pf}
\end{equation}%
for all $\beta >0$.

A comparison of the results in (\ref{raw_bs}) and (\ref{pf}) highlights the
impact of the pre-filtering on the ability of the sieve bootstrap to
accurately reproduce the sampling distribution in question. Whilst both
techniques achieve higher-order convergence, the rate of convergence of the
pre-filtered algorithm is arbitrarily close to the $O_{p}(T^{-1+\beta })$
rate achieved with simple random samples, for any pre-filtering estimate $%
\widehat{d}$ that converges almost surely to the true value of $d$ at the
appropriate rate as $T$ $\rightarrow \infty $. Clearly, the more accurate
the preliminary estimate of $d$ (i.e. the speed with which $\delta _{T}\log
T $\ approaches zero in practice) the more useful the pre-filtering, in
terms of yielding a filtered process for which the autoregressive
approximation and, ultimately, the distributional estimate, is accurate for
any given value of $T$. Given the non-parametric nature of our approach, in
the simulation exercise that follows we apply an algorithm based on a
pre-filtering value equivalent to the SPLW estimator of \cite{robinson:1995b}%
, where the estimator is constrained to lie in the stationary region. As a
corollary of \citet[Lemma
5.8]{giraitis:robinson:2003} this estimator satisfies $P(|\widehat{d}-d|\ln
T>\epsilon )=o(N^{-p})$, where $p>1/\epsilon $ and $N$, the bandwidth,
satisfies $T^{\epsilon }<N<T^{1-\epsilon }$ for some $\epsilon >0$. As such,
the almost sure limiting criterion required of the pre-filtering value holds
and the $O(T^{-1+\beta })$ convergence rate for the sieve method is
attainable.\footnote{%
The current pre-filtering value, $\widehat{d}$, has been chosen because it
has been shown to satisfy the required large deviations property. As pointed
out by a referee, $\widehat{d}$ is an early version of the possible
semi-parametric estimators of $d$, and there are more recent estimators that
have been shown to have better finite sample properties. Consistency and
asymptotic normality have been established for these latter estimators, but
the relevant limiting criterion has not, to our knowledge, been proven. It
is beyond the scope of this paper to establish the required large deviations
result for these estimators, and to undertake a comparison of the finite
sample results that would follow from different choices of such pre-filters.}%
\vspace{-12pt}

\section{Properties of persistence measures for a fractional process}

\subsection{The sample autocorrelation function}

Following \cite{hosking:1996}, we define the $k^{th}$ sample autocorrelation
coefficient as
\begin{equation}
\widehat{\rho }(k)=\frac{\textstyle\sum\nolimits_{t=1}^{T-k}(y(t)-\bar{y}%
_{T})(y(t+k)-\bar{y}_{T})}{\textstyle\sum\nolimits_{t=1}^{T}(y(t)-\bar{y}%
_{T})^{2}},  \label{phat}
\end{equation}%
where $\bar{y}_{T}=\frac{1}{T}\textstyle\sum\nolimits_{t=1}^{T}y(t)$.
Hosking's \citeyearpar{hosking:1996} summary of the asymptotic properties of
$\widehat{\rho }(k)$\ under long memory includes the following expression
for the large-sample bias:\footnote{%
Note that Hosking's symbol $\alpha $ corresponds to $1-2d$ in the notation
used here.}
\begin{equation}
Bias\left[ \widehat{\rho }(k)\right] \sim \frac{-\lambda }{d(1+2d)}\left\{
\frac{1-\rho (k)}{\gamma (0)}\right\} T^{2d-1},  \label{asy_bias}
\end{equation}%
where $\lambda =\{\sigma \kappa (1)\}^{2}\frac{\Gamma (1-2d)}{\Gamma
(d)\Gamma (1-d)}$. This is seen to be negative for all $-0.5<d<0.5$. In
addition, for $0.25<d<0.5$ the normalized quantity $\frac{T^{1-2d}}{(1-\rho
(k))}(\widehat{\rho }(k)-\rho (k))$, $k=0,1,\ldots ,T-1$, converges in
distribution to the `modified Rosenblatt', with cumulants as documented in
\citet[Table
2]{hosking:1996}. Most notably, the mean of this limiting distribution is
shown to be both substantially less than zero for all $d>0.25$, and larger
in magnitude than the standard deviation for $d>0.35$. Hence, in cases where
the true persistence in the process is high, it is to be anticipated that
the sample autocorrelation function will substantially underestimate the
extent of this persistence. Further, in this case, an approximating normal
distribution is inappropriate in terms of capturing sampling variation in
the estimated autocorrelation coefficients.

The definition in \eqref{phat} is, of course, only one of several
asymptotically equivalent estimators of $\rho (k)$. \cite{lee:ko:2009}
instead consider
\begin{equation}
r(k)=\frac{C(k)}{C(0)}=\frac{\frac{1}{T-k}\textstyle\sum%
\nolimits_{t=1}^{T-k}(y(t)-\overline{y}_{[1:T-k]})(y(t+k)-\overline{y}%
_{[k+1:T]})}{\frac{1}{T}\textstyle\sum\nolimits_{t=1}^{T}(y(t)-\overline{y}%
_{T})^{2}},  \label{LK}
\end{equation}%
where $\overline{y}_{[1:T-k]}=\textstyle\sum\nolimits_{t=1}^{T-k}y(t)/(T-k)$
and $\overline{y}_{[k+1:T]}=\textstyle\sum\nolimits_{t=k+1}^{T}y(t)/(T-k)$;
and proceed to derive a closed-form expression for the bias of $r(1)$ based
on the much earlier work of \cite{marriott:pope:1954}, in which, up to $%
O(T^{-1}),$ the expected value of the $k^{th}$-order sample autocorrelation
coefficient is shown to be
\begin{equation}
E(r(k))=\frac{E\left[ C(k)\right] }{E\left[ C(0)\right] }\left[ 1-\frac{%
cov[C(k),C(0)]}{E\left[ C(k)\right] E\left[ C(0)\right] }+\frac{var[C(0)]}{%
E^{2}[C(0)]}\right] .  \label{exp_phat}
\end{equation}%
\cite{newbold:agiakloglou:1993} earlier evaluated (\ref{exp_phat}) under a
Gaussian fractional noise process (produced by setting $\kappa (z)=1$ in \ref%
{Wold}). Their results demonstrate a distinct negative bias in $r(k)$ for
all values of $k$ considered, and are consistent with the expectation --
given the asymptotic results of \cite{hosking:1996} -- that this bias is
more pronounced the larger is $d$. \citeauthor{newbold:agiakloglou:1993}
also find the bias to be even more pronounced in the empirically relevant
case considered here, in which the sample mean is used in the calculation of
the sample autocorrelations, compared to the artificial scenario in which
the mean is assumed known.

\cite{lee:ko:2009} use the expression in (\ref{exp_phat}) to produce a
closed-form \textquotedblleft exact to $O(T^{-1})$\textquotedblright\
representation of the bias of $r(1)$ in terms of the true $\rho (1)\ldots
\rho (T-1)$, and plot the ratio of this `first-order' bias to the $%
O(T^{2d-1})$ asymptotic bias in (\ref{asy_bias}) for various values of $T$
and $d$ under the assumption of fractional noise. In this case the
asymptotic measure is shown to underestimate the first-order measure for any
$d>0$, with the extent of this underestimation increasing rapidly with $d$. %
\citeauthor{lee:ko:2009} use their expression, evaluated at a preliminary
estimate of $d$ (upon which this expression naturally depends) to bias
correct $r(1)$ and so produce a simple \textquotedblleft
bias-adjusted\textquotedblright\ method of moments estimator of $d$. They do
not, however, explicitly examine the sampling properties of the
bias-corrected estimator of $\rho (1)$ itself.

\subsection{The impulse response function}

As noted above, our focus is on bias-adjusting semi-parametric estimates of
the $k^{th}$ impulse response coefficient $\psi (k)$ defined in (\ref{Wold}%
). The basic semi-parametric estimation procedure involves fitting an
autoregressive model of order $h$ (to be determined) to $y(t)$ and
inverting, to produce $\widehat{\psi }(k)$ as the $k^{th}$ term in the
expansion

\begin{equation}
\widehat{\psi }(z)=\widehat{\Phi }_{h}^{-1}(z)=\textstyle\sum\limits_{k=1}^{%
\infty }\widehat{\psi }(k)z^{k},  \label{irf}
\end{equation}%
where $\widehat{\Phi }_{h}(z)=1+\hat{\phi}_{h}(1)z+\hat{\phi}%
_{h}(2)z^{2}+....+\hat{\phi}_{h}(h)z^{h},\,$\ and the $\phi _{h}(j)$, $%
j=1,2,...,h$ are estimated as described in Section \ref{sieve}. As
documented in \cite{baillie:kapetanios:2013}, use of this approach in the
long memory setting yields more accurate estimates of the true impulse
response coefficients than do certain mis-specified parametric methods, and
may even be competitive with correctly specified parametric methods for some
parameter combinations. However, as we also document below, a marked
negative bias is still a characteristic of these semi-parametric estimates.
\cite{baillie:kapetanios:2013} produce a bias-adjusted estimate of the IRF
by using the bootstrap technique of \cite{kilian:1998} to bias-adjust the
estimated autoregressive coefficients prior to inverting to them to produce
the $\widehat{\psi }(k)$. In contrast, we bias correct the $\widehat{\psi }%
(k)$ directly, as described in detail in the next section. Pre-empting our
results, we find that the use of the pre-filtered sieve produces
bias-adjusted statistics that are very similar to those produced by our
adaptation of the Kilian method, but with the pre-filtering method yielding
more accuracy when both the sample is small and the level of persistence in
the data is high.\vspace{-12pt}

\section{Simulation Exercise\label{sim}}

In this section we examine the performance of the raw and pre-filtered sieve
algorithms via a simulation experiment. Specifically, we investigate the
finite sample accuracy of both forms of bootstrap-based bias-adjusted
estimates of the autocorrelation and impulse response coefficients,
documenting the remaining bias and root mean squared error across Monte
Carlo replications, as well as plotting selected sampling distributions.
Corresponding results for the unadjusted statistics are also included, in
order to demonstrate the extent of the improvement yielded by the
bias-adjustment techniques. We also consider the accuracy with which the
bootstrap algorithms reproduce the `true' (Monte Carlo) sampling
distribution of the unadjusted persistence statistics, in selected cases, as
it is these bootstrap distributions that underlie the subsequent
bias-adjustment.

\subsection{Simulation design and computational details\label{comp}}

Data are simulated from a zero mean Gaussian ARFIMA$(1,d,0)$ process,
\begin{equation}
(1-L)^{d}\Phi (z)y(t)=\varepsilon (t)\,,\ 0<d<0.5\,,  \label{arfima}
\end{equation}%
with $\Phi (z)=1-\phi z$ being the operator for a stationary AR(1) component
and $\varepsilon (t)$ is zero-mean Gaussian white noise. The process in (\ref%
{arfima}) is simulated $R=1000$ times for $d=\left\{ 0.2,0.4\right\} $, $%
\phi =\left\{ 0.6,0.9\right\} $, and sample sizes $T=100$ and $500$ via
Levinson recursion applied to the autocovariance function of the desired
ARFIMA$(1,d,0)$ process and the generated pseudo-random $\varepsilon (t)$
\citep[see, for
instance,][\S5.2]{brockwell:davis:1991}. The autocovariance function for
given $T$, $\phi $ and $d$ is calculated using Sowell's %
\citeyearpar{sowell:1992} algorithm as modified by \cite{doornik:ooms:2003}.
Parameter settings are chosen that yield, respectively, moderate and large
bias in both the estimated IRF and the estimated ACF.

For each realization $r$ of the process we compute the relevant scalar
statistic, $s_{T,r}$, plus $B=1000$ bootstrap estimates $s_{T,r(b)}^{\ast }$%
, constructed using $b=1,\ldots ,B$ bootstrap re-samples obtained via the
sieve algorithm. Each realized value $s_{T,r}$ thus has associated with it a
`bootstrap distribution'\ based on the $B$ bootstrap resamples $%
s_{T,r(b)}^{\ast }$, $b=1,\ldots ,B,$ with each such distribution serving as
an estimate of the sampling distribution of $s_{T}$. In order to compare the
$R$ bootstrap distributions with the finite sample distribution estimated
from the Monte Carlo draws, we first sort the $B$ bootstrap draws for each
MC replication into ascending order, then average these ordered bootstrap
values across the Monte Carlo draws. The $B$ averaged draws are then used to
produce a kernel density estimate, which we refer to as the `average'
bootstrap distribution.

Our focus is on two types of statistic: $s_{T}=\widehat{\rho }(k)$, computed
as per (\ref{phat}), and $s_{T}=\widehat{\psi }(k)$, computed as per (\ref%
{irf}), for $k=1,2,...99$; and on using the sieve bootstrap techniques to
bias adjust each. Specifically, for any given realization $r$, the bootstrap
distribution (computed from the $B$ bootstrap resamples) is used to produce
an estimate of $E(s_{T})$, $\widehat{E}(s_{T})$; and a bias-adjusted
statistic,
\begin{equation}
s_{T,r}^{(BA)}=s_{T,r}-\widehat{bias}(s_{T}),  \label{ba_stat}
\end{equation}%
thereby constructed, where
\begin{equation}
\widehat{bias}(s_{T})=\widehat{E}(s_{T})-s_{ref},  \label{bias}
\end{equation}%
and $s_{ref}$ denotes the appropriate reference value to be used in the
definition of the bias, the construction of which is elaborated on below.
The sampling distribution of this statistic is then estimated from the $R$
Monte Carlo draws using kernel density methods and the finite sample
performance of the statistic as an estimator of the true parameter
summarized via its bias and root mean square error (RMSE). The two different
forms of sieve bootstrap (raw and pre-filtered) produce a different estimate
$\widehat{E}(s_{T})$ and, as will be made clear below, a different value for
$s_{ref}.$ Hence, for both reasons, each algorithm produces a different bias
estimate in (\ref{bias}), and a different bias-adjusted statistic in (\ref%
{ba_stat}).

With regard to specifying the order of the autoregressive approximation used
in the sieve, we begin by specifying, as is common practice %
\citep[][\S3]{politis:2003}, $h=\hat{h}_{T}=\mathrm{argmin}_{h=0,1,\ldots
,M_{T}}\left( \ln (\hat{\sigma}_{h}^{2})+2h/T\right) $, where $\hat{\sigma}%
_{h}^{2}$ denotes the residual mean square obtained from an $AR(h)$ model
and $M_{T}=[(\ln T)^{2}]$. This procedure (order selection via Akaike's %
\citeyearpar{akaike:1973} information criterion, or AIC) is asymptotically
efficient in the sense of being equivalent to minimizing Shibata's %
\citeyearpar{shibata:1980} figure of merit. For comparison we also consider $%
h=h_{T}=[(\ln T)^{2}],$ this being the fixed (for given $T$) value of $h_{T}$
used by \cite{baillie:kapetanios:2013}.

Note that in the case of the IRF the alternative values for $h$ are relevant
not only in defining the order of the fitted autoregression in the sieve,
and hence the bootstrap `data generating process' from which the reference
values used in the bias calculations (for both the IRF and the ACF) are
backed out; $h$ also defines the order of the autoregression used to obtain
the sample impulse response coefficients themselves (i.e., the actual
statistics being bootstrapped and bias adjusted). Accordingly, when
bootstrapping the IRF we set the order of the sieve approximation to be
consistent with the order of the autoregression used to produce the IRF\
estimator being examined. That is, when $\widehat{\psi }(k)$\ is produced
via an autoregression with fixed order $h_{T}$,\ the order of the sieve used
in the bootstrap, whether raw or pre-filtered, is also set to $h_{T}$.\
Similarly, when $\widehat{\psi }(k)$\ is produced via an autoregression with
order selected by AIC, the order of the sieve used in the bootstrap is also
selected by AIC. When using the raw sieve this naturally means that the
sieve and estimating autoregression are exactly the same. However, this last
is not the case when we switch to the pre-filtered method.

In order to render the bootstrap estimate of the bias a valid representation
of the true but unknown bias, the reference value, $s_{ref}$,\ used in the
bias computation for each of the two measures, is defined in a way that is
consistent with the method used to generate the bootstrap samples.
Accordingly,\ the reference value for bias adjustment in the case of the raw
sieve algorithm is that implied by the $AR(h)$ sieve (where $h$ may be $\hat{%
h}_{T},$ $h_{T},$ or any other value that increases at the appropriate rate
in $T$) fitted to the raw data $y(t)$ (rather than the pre-filtered series $%
\widehat{w}(t)$) in Step 2.\ref{s1} in Section \ref{pf_alg}. Denoting this
by
\begin{equation}
\overline{\Phi }_{h}\left( z\right) =1+\bar{\phi}_{h}(1)z+\cdots +\bar{\phi}%
_{h}(h)z^{h},  \label{ar_coeff}
\end{equation}%
the reference IRF appropriate to the raw sieve is accordingly produced by
the inversion of $\overline{\Phi }_{h}\left( z\right) $, whilst the
corresponding reference ACF follows via the Yule-Walker equations.

The pre-filtered sieve method, on the other hand, implies an ARFIMA$(h,\hat{d%
},0)$ bootstrap model, with $\hat{d}$ the pre-filtering fractional
integration parameter, and autoregressive\ coefficients $\hat{\phi}%
_{h}(1),\ldots ,\hat{\phi}_{h}(h)$ produced by fitting an $AR(h)$ to the
filtered data $(1-z)^{\hat{d}}y(t)$. The reference IRF, $\tilde{\psi}(k)$,
is therefore now obtained by inverting the implied ARFIMA$(h,\hat{d},0)$
polynomial; i.e., $\tilde{\psi}(z)=\widehat{\Phi }_{h}^{-1}(z)(1-z)^{-\hat{d}%
}$ (cf. \ref{irf}); while the reference ACF $\tilde{\rho}(k)$ is calculated
by applying the \citeauthor{sowell:1992}/\citeauthor{doornik:ooms:2003}
algorithm to the implied ARFIMA$(h,\hat{d},0)$ model. We note here that the
restriction of the pre-filtering SPLW estimate to the stationary region is
essential at this point.

Finally, we note that in order to produce bias-corrected estimates of $\rho
(k)$ that necessarily lie between minus one and one we perform our bias
correction in terms of the so-called \textquotedblleft Fisher-$z$%
\textquotedblright\ transformation, which maps from any $r\in \left(
-1,1\right) $ to $\zeta \in \mathbb{R}$ via $\zeta =\frac{1}{2}\ln \left(
\frac{1+r}{1-r}\right) =\func{artanh}\left( r\right) $. That is, while the
statistic of interest is still $\widehat{\rho }(k)$, the bootstrap bias
correction (and therefore the bootstrapping itself) is done in terms of $%
\zeta \left( \widehat{\rho }(k)\right) $, with the bootstrap-bias-adjusted
estimate of $\rho (\cdot )$ produced via the reverse mapping
\begin{equation}
r=\frac{e^{2\zeta }-1}{e^{2\zeta }+1}=\tanh \left( \zeta \right) .
\label{ifz}
\end{equation}%
It is the reverse mapping $\zeta \rightarrow r$ that ensures that the
bias-corrected result is within $\left( -1,1\right) $.

For the ACF we also plot results for two additional bias-corrected
quantities: one based on the subtraction of (\ref{asy_bias}) from $\widehat{%
\rho }(k)$, with all unknown parameters in (\ref{asy_bias}) assigned their
true values from the data generating process; and the second (for the case
of $k=1$ only) based on the subtraction of an estimate of the \cite%
{lee:ko:2009} $O\left( T^{-1}\right) $ bias expression from $\widehat{\rho }%
(1)$.\footnote{%
The statistics $r(k)$ and $\widehat{\rho }(k)$ are such that $r(k)=\widehat{%
\rho }(k)+O(T^{-1}).$ Hence the $O(T^{-1})$ bias result for $r(k)$ produced
by Lee and Ko applies to $\widehat{\rho }(k)$ also.} The former (theoretical
asymptotic-bias-adjusted ACF) is denoted by $\widehat{\rho }^{(ASY)}(k)$;
the latter (estimated $O\left( T^{-1}\right) $-bias-adjusted $\widehat{\rho }%
(1)$) by $\widehat{\rho }^{(LK)}(1)$. The Lee and Ko bias is estimated by
replacing the unknown population autocorrelations in their bias formula by
the $\tilde{\rho}(k)$'s implied by an $AR(h_{T})$ fitted to the unfiltered
data. In other words, the Lee and Ko bias is calculated using the reference\
ACF corresponding to the raw sieve as described above, with $h=h_{T}=\left[
(\ln T)^{2}\right] $.

For interest, we also present results based on a modification of the method
of \cite{kilian:1998} for bias adjusting the IRF. In brief, our version of
Kilian's method involves using the raw sieve bootstrap to bias correct the
autoregressive coefficients in (\ref{ar_coeff}), then inverting the
resulting bias-adjusted polynomial to produce an estimate of the IRF. Our
approach differs slightly from that of Kilian in that: firstly, our
estimates of the autoregressive coefficients are obtained via the Burg
algorithm rather than OLS, and hence the issue of potentially non-stationary
coefficient estimates does not arise; secondly, stationarity is preserved
after bias-correction by applying the Schur-Cohn stability test and
reflecting any zeroes found to be outside $\{|z|=1\}$ back inside the unit
circle, rather than by iteratively shrinking the bias-corrected
autoregressive operator.

\subsection{Simulation Results\label{simres}}

Due to space considerations, we present here selected results for the sample
IRF and ACF based on $T=500$ only. Corresponding results for $T=100$ can be
found in the \href{http://users.monash.edu.au/~gmartin/Grose_Martin_Poskitt_on_line_appendix.pdf}%
{Supplementary Appendix}. As would be expected, the performance of the
bootstrap-based methods improves with an increase in the sample size.
However, we explicitly discuss the $T=100$ results in the text only when
they differ qualitatively from those for $T=500$.

\subsubsection{Bias correction of the sample IRF\label{results_irf}}

Panels (i) to (v) in each figure plot the Monte Carlo distribution of the
unadjusted\ statistic $\widehat{\psi }(k)$; the Monte Carlo distribution of
the bootstrap bias-adjusted statistic $\widehat{\psi }^{(BA)}(k)$; and the
average bootstrap estimate of the distribution of $\widehat{\psi }(k)$.
(These are indicated by the legend entries \textquotedblleft
MC\textquotedblright , \textquotedblleft MC-BA\textquotedblright\ and
\textquotedblleft BS-av\textquotedblright\ respectively). The vertical
dotted line in each panel indicates the position of the true value of $\psi
(k)$ for each $k=1,3,6,9,12$. Panel (vi) plots, for lags $k=1,2,...,99$, the
true IRF $\psi (k)$ (based on the parameters of the true data generating
process); the mean of the Monte Carlo distribution of $\widehat{\psi }(k)$;
and the mean of the Monte Carlo distribution of $\widehat{\psi }^{(BA)}(k)$
(designated \textquotedblleft True $\psi $\textquotedblright ,
\textquotedblleft $\overline{MC}$\textquotedblright , and \textquotedblleft $%
\overline{MC\text{-}BA}$\textquotedblright\ respectively).

\begin{figure}[tbph]
\centering%
\includegraphics[trim=10mm 70mm 10mm
80mm,width=6in]{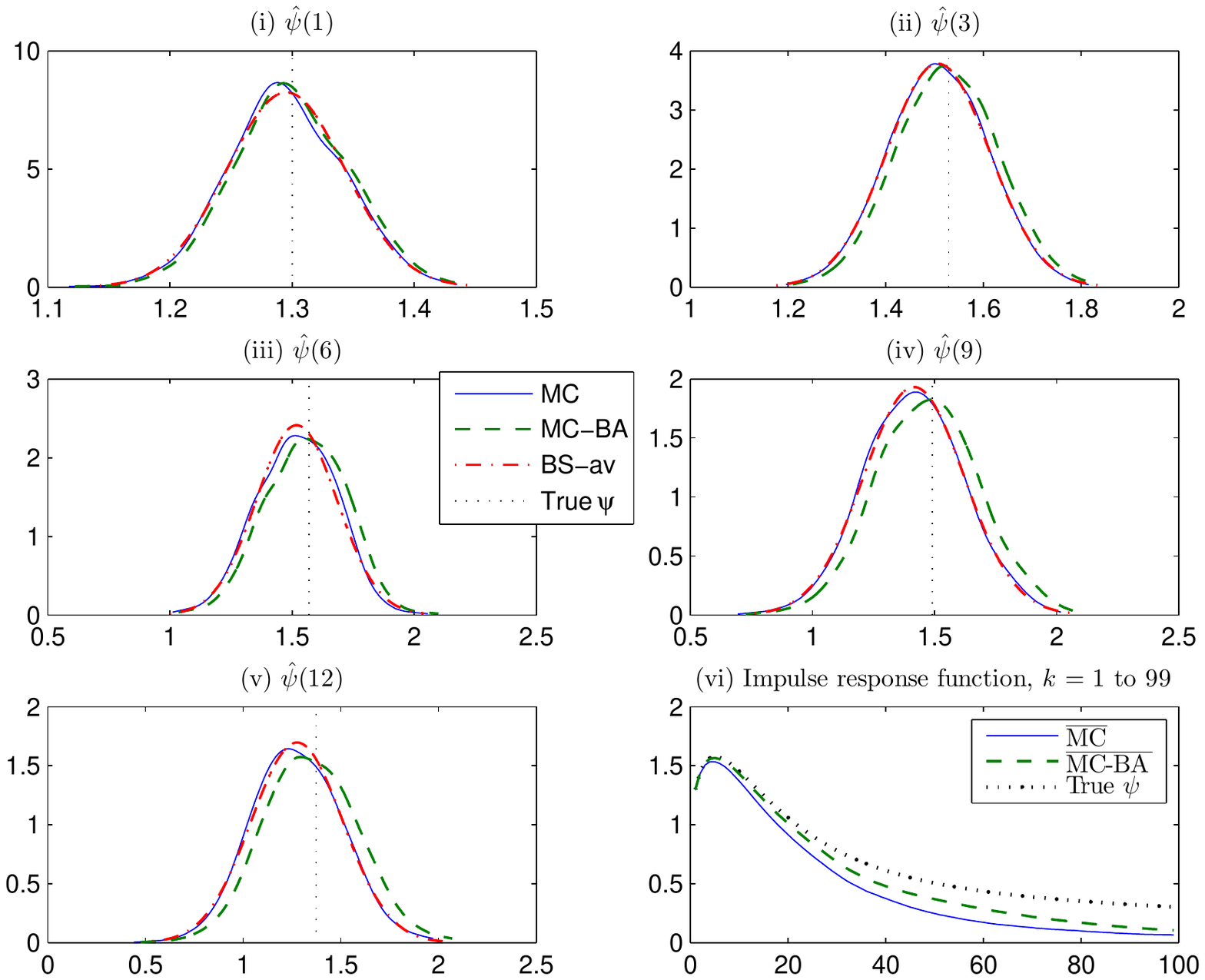}
\caption{Bias correction of the sample IRF using the \textit{raw sieve}
bootstrap. \protect\smallskip \newline
\textit{True process: }ARFIMA$(1,d,0)$; $T=500$; $d=0.4$; $\protect\phi =0.9$%
. \protect\smallskip \newline
\textit{Key for Panels (1) to (v): }MC\textquotedblright : Monte Carlo
distribution of the unadjusted\ statistic $\protect\widehat{\protect\psi }%
(k) $; \textquotedblleft MC-BA\textquotedblright : Monte Carlo distribution
of the bootstrap bias-adjusted statistic $\protect\widehat{\protect\psi }%
^{(BA)}(k)$; \textquotedblleft BS-av\textquotedblright : the averaged
bootstrap estimate of the distribution of $\protect\widehat{\protect\psi }%
(k) $. \textit{Key for Panel (vi):} \textquotedblleft $\overline{MC}$%
\textquotedblright : mean of the Monte Carlo distribution of $\protect%
\widehat{\protect\psi }(k)$; \textquotedblleft $\overline{MC\text{-}BA}$%
\textquotedblright : mean of the Monte Carlo distribution of $\protect%
\widehat{\protect\psi }^{(BA)}(k)$. The true value of $\protect\psi (k)$ is
indicated by the use of small dots in all panels.}
\label{fig2}
\end{figure}

\begin{figure}[tbph]
\centering%
\includegraphics[trim=10mm 70mm 10mm
80mm,width=6in]{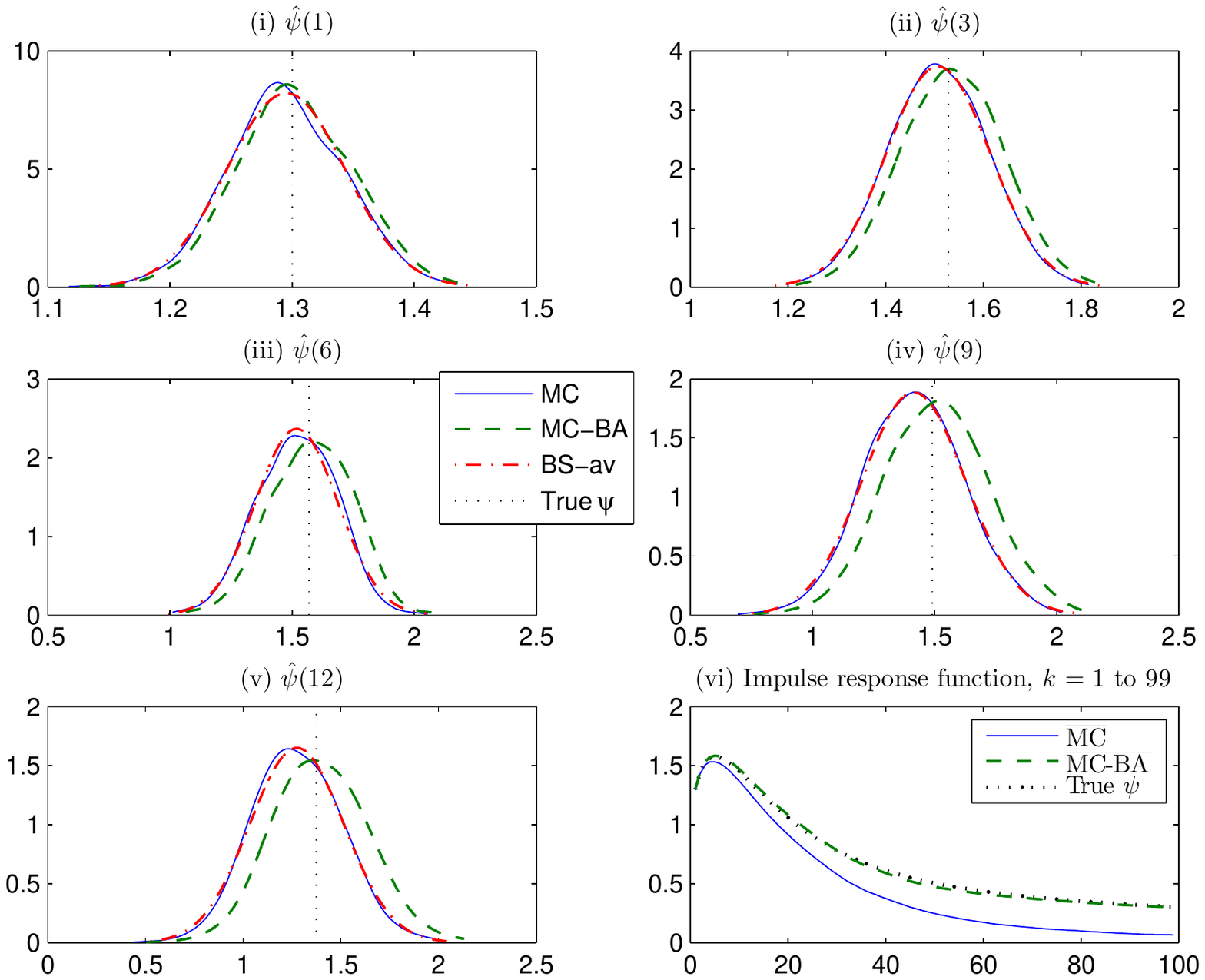}
\caption{Bias correction of the sample IRF using the \textit{pre-filtered
sieve} bootstrap, based on the \textit{true value of }$d$ \textit{as the
pre-filter}.\protect\smallskip \newline
\textit{True process: }ARFIMA$(1,d,0)$; $T=500$; $d=0.4$; $\protect\phi =0.9$%
. \protect\smallskip \newline
\textit{Key for Panels (1) to (v): }MC\textquotedblright : Monte Carlo
distribution of the unadjusted\ statistic $\protect\widehat{\protect\psi }%
(k) $; \textquotedblleft MC-BA\textquotedblright : Monte Carlo distribution
of the bootstrap bias-adjusted statistic $\protect\widehat{\protect\psi }%
^{(BA)}(k)$; \textquotedblleft BS-av\textquotedblright : the averaged
bootstrap estimate of the distribution of $\protect\widehat{\protect\psi }%
(k) $. \textit{Key for Panel (vi):} \textquotedblleft $\overline{MC}$%
\textquotedblright : mean of the Monte Carlo distribution of $\protect%
\widehat{\protect\psi }(k)$; \textquotedblleft $\overline{MC\text{-}BA}$%
\textquotedblright : mean of the Monte Carlo distribution of $\protect%
\widehat{\protect\psi }^{(BA)}(k)$. The true value of $\protect\psi (k)$ is
indicated by the use of small dots in all panels.}
\label{fig4}
\end{figure}

\begin{figure}[tbph]
\includegraphics[trim=10mm 70mm 10mm
80mm,width=6in]{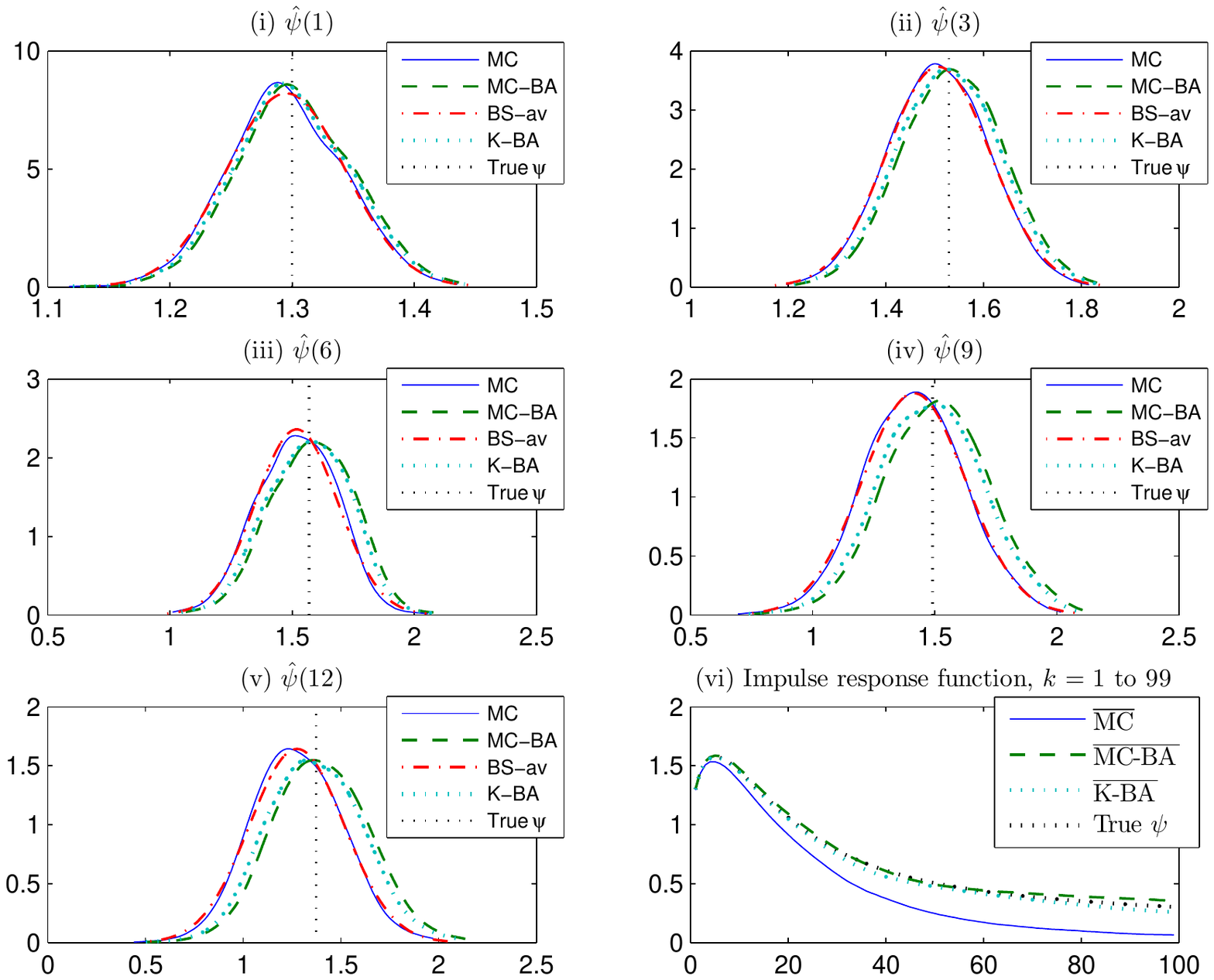}
\caption{Bias correction of the sample IRF using the \textit{pre-filtered
sieve bootstrap}, based on the \textit{SPLW estimate of} $d$ \textit{as the
pre-filter.}\protect\smallskip \newline
\textit{True process:} ARFIMA$(1,d,0)$; $T=500$; $d=0.4$; $\protect\phi =0.9$%
. \protect\smallskip \newline
\textit{Key for Panels (1) to (v): }\textquotedblleft MC\textquotedblright :
Monte Carlo distribution of the unadjusted\ statistic $\protect\widehat{%
\protect\psi }(k)$; \textquotedblleft MC-BA\textquotedblright : Monte Carlo
distribution of the bootstrap bias-adjusted statistic $\protect\widehat{%
\protect\psi }^{(BA)}(k)$; \textquotedblleft BS-av\textquotedblright :\
average bootstrap estimate of the distribution of $\protect\widehat{\protect%
\psi }(k)$, and \textquotedblleft K-BA\textquotedblright :\ Monte Carlo
distribution of the bias-adjusted statistic $\protect\widehat{\protect\psi }%
^{(K)}(k)$ produced using Kilian's approach. \textit{Key for Panel (vi): }%
\textquotedblleft $\overline{MC}$\textquotedblright : mean of the Monte
Carlo distribution of $\protect\widehat{\protect\psi }(k)$;
\textquotedblleft $\overline{MC\text{-}BA}$\textquotedblright : mean of the
Monte Carlo distribution of $\protect\widehat{\protect\psi }^{(BA)}(k)$;
\textquotedblleft $\overline{K\text{-}BA}$\textquotedblright : mean of the
Monte Carlo distribution of $\protect\widehat{\protect\psi }^{(K)}(k)$. The
true value of $\protect\psi (k)$ is indicated by the use of small dots in
all panels.}
\label{fig6}
\end{figure}

Figure \ref{fig2} displays the distributional results as listed above for $%
d=0.4$ and $\phi =0.9$, where the bias adjustment occurs via the raw sieve
algorithm, and $h$ $=\left[ (\ln T)^{2}\right] $. Bias and RMSE results for
both choices of $h$, and for all combinations of $d=0.2,0.4$ and $\phi
=0.6,0.9$, are presented in Table \ref{newtbl1}.

The first thing to note from Figure \ref{fig2}, and something that will be a
feature of all graphs included both in the body of the paper and in the
appendix, is the accuracy with which the sieve (and, to an even greater
extent, the pre-filtered sieve) technique reproduces the true sampling
distribution of the statistic to be bias adjusted. This result (including
the overall improvement in fit that will be seen to be yielded by the
pre-filtering) is consistent with the supporting theoretical convergence
results cited in \S \ref{converg}, and provides further support for using
the bootstrap-based estimate of the sampling distribution as a basis for
estimating the bias of any given statistic, and bias adjusting subsequently.
As is clear from Figure \ref{fig2}, the negative finite sample bias of $%
\widehat{\psi }^{(BA)}(k)$ documented in \cite{baillie:kapetanios:2013} is
in evidence here, for all lags $k$, with the magnitude of the bias
increasing with $k$ up to approximately $k=40$, then leveling out thereafter
to a fairly constant value. The bootstrap-based bias adjustment is seen to
produce a very accurate bias-adjusted estimator for low values of $k$, and
to continue to yield improvements over the unadjusted statistic for all
values of $k$ considered.

In Figure \ref{fig4} we plot the corresponding results based on the
pre-filtered bootstrap technique with the \textit{true} value of $d$ used in
the pre-filtering. These results provide resounding proof-of-concept support
for the pre-filtering technique, with the bootstrap-based bias-adjusted
estimator\ seen to be very accurate, indeed to have a mean value (across
Monte Carlo replications) that is almost visually indistinguishable from the
true $\psi (k)$\ for all values of $k$\ considered.

An empirically feasible version of the pre-filtering technique requires the
use of an estimate of $d$ as the pre-filtering value, with the constrained
SPLW estimator of \cite{robinson:1995b} used for this purpose. As
highlighted in Figure \ref{fig6}, we observe excellent bias correction for
the lower values of $k$, with the sampling distributions of the adjusted
statistic (MC-BA in the graphs) located quite precisely with respect to the
true value of the IRF in each case, and with very little cost in terms of
additional dispersion. Note that, although we haven't included this figure
here, for the medium persistence design ($d=0.2,\phi =0.6$) the SPLW-based
pre-filtering technique does tend to very slightly `over-correct' for the
longer lag lengths ($k>45$), where the method that exploits the true value
of $d$ as the pre-filter does not. Overall, however, very little accuracy is
lost via the substitution of $\widehat{d}$ for $d$, with the bias-adjusted
estimator remaining remarkably accurate.

Figure \ref{fig6} also includes the Monte Carlo estimate of the distribution
of the bias-adjusted estimator produced using Kilian's %
\citeyearpar{kilian:1998} method, $\widehat{\psi }^{(K)}(k),$ modified as
described in \S \ref{comp}. We see that the Kilian-based method (denoted by
K-BA in the figure) yields very similar accuracy to the pre-filtered
bootstrap technique for $T=500$. However, as will be noted from the
corresponding figure for $T=100$ included in the appendix, the pre-filtering
method is more successful in correcting the more substantial bias that
obtains in this case; although both methods are certainly superior to the
raw sieve method. Further results (available on request) confirm the general
accordance between the pre-filtered sieve and Kilian approaches.

\afterpage{
\begin{landscape}
\setlength{\oddsidemargin}{-0.4in}
\setlength{\textwidth}{6.8in}

\begin{table}[tbp] \centering
\caption{Bias and root mean squared error (RMSE) of estimators of selected impulse
response coefficients, for $T=500$.\\
Results for the unadjusted and both forms of bootstrap-based bias-adjusted
estimators are documented.}\label{newtbl1}

\begin{tabular}{ccccccccccccccccc}
&  &  &  &  &  &  &  &  &  &  &  &  &  &  &  &  \\[-2.5ex]
&  &  & \multicolumn{4}{c}{$\widehat{\psi }(k)$} &  & \multicolumn{4}{c}{$\widehat{\psi }^{(BA)}(k)$ (raw sieve)} &  & \multicolumn{4}{c}{$\widehat{\psi }^{(BA)}(k)$ (pre-filtered sieve)} \\
\cline{4-7}\cline{9-12}\cline{14-17}
&  &  &  &  &  &  &  &  &  &  &  &  &  &  &  &  \\[-1.5ex]
&  &  & $k=1$ & $k=6$ & $k=12$ & av. &  & $k=1$ & $k=6$ & $k=12$ & av. &  & $k=1$ & $k=6$ & $k=12$ & av. \\[2ex]
\multicolumn{17}{c}{Panel A: $T=500;$ $h$ based on AIC selection} \\[0.5ex] \hline
&  &  &  &  &  &  &  &  &  &  &  &  &  &  &  &  \\[-1ex]
$d$ & $\phi $ & \multicolumn{15}{c}{Bias} \\[1ex]
0.2 & 0.6 &  & -0.0111 & -0.0026 & -0.0310 & -0.0126 &  & -0.0057 & 0.0056 &
-0.0282 & -0.0063 &  & -0.0190 & -0.0140 & 0.0414 & -0.0032 \\
& 0.9 &  & -0.0020 & -0.0614 & -0.0610 & -0.0447 &  & 0.0023 & -0.0283 &
-0.0109 & -0.0151 &  & 0.0027 & -0.0106 & -0.0224 & -0.0091 \\
0.4 & 0.6 &  & -0.0089 & -0.0011 & -0.0219 & -0.0092 &  & -0.0043 & 0.0196 &
0.0002 & 0.0075 &  & 0.0016 & -0.0278 & 0.0050 & -0.0108 \\
& 0.9 &  & 0.0004 & -0.0944 & -0.1617 & -0.0841 &  & 0.0048 & -0.0521 &
-0.0728 & -0.0399 &  & 0.0007 & 0.0184 & -0.0097 & 0.0056 \\
&  &  &  &  &  &  &  &  &  &  &  &  &  &  &  &  \\[-1ex]
&  & \multicolumn{15}{c}{RMSE} \\[1ex]
0.2 & 0.6 &  & 0.0453 & 0.0612 & 0.0532 & 0.0548 &  & 0.0438 & 0.0634 & 0.055
& 0.0561 &  & 0.0545 & 0.0543 & 0.0575 & 0.0582 \\
& 0.9 &  & 0.0472 & 0.1190 & 0.1304 & 0.1027 &  & 0.0471 & 0.1064 & 0.1246 &
0.0956 &  & 0.0443 & 0.1223 & 0.1388 & 0.1055 \\
0.4 & 0.6 &  & 0.0525 & 0.0963 & 0.0995 & 0.0846 &  & 0.0513 & 0.1017 &
0.1048 & 0.088 &  & 0.0442 & 0.0991 & 0.0885 & 0.0817 \\
& 0.9 &  & 0.0464 & 0.1874 & 0.2524 & 0.1645 &  & 0.0467 & 0.1692 & 0.2148 &
0.1465 &  & 0.0438 & 0.1554 & 0.2280 & 0.1442 \\
&  &  &  &  &  &  &  &  &  &  &  &  &  &  &  &  \\
\multicolumn{17}{c}{Panel B: $T=500;$ $h=(\ln T)^{2}$} \\[0.5ex] \hline
&  &  &  &  &  &  &  &  &  &  &  &  &  &  &  &  \\[-1ex]
$d$ & $\phi $ & \multicolumn{15}{c}{Bias} \\[1ex]
0.2 & 0.6 &  & -0.0030 & -0.0130 & -0.0139 & -0.0105 &  & 0.0009 & -0.0031 &
-0.0032 & -0.0018 &  & 0.0034 & 0.0058 & 0.0085 & 0.0061 \\
& 0.9 &  & -0.0036 & -0.0296 & -0.0469 & -0.0266 &  & 0.0010 & -0.0040 &
-0.0073 & -0.0030 &  & 0.0039 & 0.0153 & 0.0280 & 0.0160 \\
0.4 & 0.6 &  & -0.0043 & -0.0245 & -0.0324 & -0.0208 &  & 0.0001 & -0.0078 &
-0.0108 & -0.0061 &  & 0.0029 & 0.0061 & 0.0100 & 0.0067 \\
& 0.9 &  & -0.0050 & -0.0512 & -0.0969 & -0.0499 &  & 0.0001 & -0.0128 &
-0.0265 & -0.0122 &  & 0.0028 & 0.0122 & 0.0256 & 0.0137 \\
&  &  &  &  &  &  &  &  &  &  &  &  &  &  &  &  \\[-1ex]
&  & \multicolumn{15}{c}{RMSE} \\[1ex]
0.2 & 0.6 &  & 0.0465 & 0.0725 & 0.0726 & 0.0664 &  & 0.0465 & 0.0727 &
0.0737 & 0.0668 &  & 0.0466 & 0.0737 & 0.0752 & 0.0677 \\
& 0.9 &  & 0.0465 & 0.1235 & 0.1558 & 0.1112 &  & 0.0464 & 0.1224 & 0.1548 &
0.1105 &  & 0.0466 & 0.1247 & 0.1603 & 0.1131 \\
0.4 & 0.6 &  & 0.0466 & 0.0964 & 0.1051 & 0.0859 &  & 0.0465 & 0.0955 &
0.1044 & 0.0854 &  & 0.0466 & 0.0964 & 0.1060 & 0.0862 \\
& 0.9 &  & 0.0466 & 0.1684 & 0.2478 & 0.1559 &  & 0.0464 & 0.1642 & 0.2392 &
0.1517 &  & 0.0466 & 0.1658 & 0.2432 & 0.1536 \\
&  &  &  &  &  &  &  &  &  &  &  &  &  &  &  &
\end{tabular}
\end{table}
\end{landscape}}

These selected graphical results are supplemented by the bias and RMSE
results presented in Table \ref{newtbl1}, in which Monte Carlo estimates of
these quantities for $d=0.2,0.4$; and $\phi =0.6,0.9$ are recorded for the
unadjusted, raw sieve bias-adjusted, and pre-filtered sieve bias-adjusted
statistics. Panel B reports results based on $h=\left[ (\ln T)^{2}\right] $
(also underlying the figures above), whilst results based on $h$ selected by
AIC are documented in Panel A. Results are reported for $k=1$, $6$ and $12$,
with the relevant average over $k=1,3,6,9,12$ also recorded in the column
headed `av'. We have not reported numerical results for the pre-filtering
method based on the true $d$. Once again, the corresponding results for $%
T=100$ are tabulated in the appendix.

Beginning with the results for the `long AR'\ ($h=\left[ (\ln T)^{2}\right] $%
)-based estimator and bootstrap (Panel B), we find that, relative to the
unadjusted estimator, the bias-adjusted estimator based on the raw sieve is
invariably superior in terms of bias for all values of $d$, $\phi $\ and $k$
here considered. Indeed, we see that use of the raw sieve to bias adjust
results in across-the-board bias reductions, essentially to zero. The RMSE,
however, is virtually identical to that of the unadjusted estimator,
indicating the increased dispersion that inevitably accompanies bias
correction based on an estimated measure of the bias. Whilst the performance
of the pre-filtered sieve algorithm for the relatively low lag values
documented in the table is slightly mixed relative to the raw method, it
generally results in an improvement relative to the unadjusted statistic.
The pre-filtered method is evidently most advantageous relative to the raw
as the lag length $k$\ increases, with the former producing a vast reduction
in bias overall, relative to the latter, when an extended spectrum of values
for $k$\ is considered, as the graphical results recorded in Figures \ref%
{fig2} and \ref{fig6} highlight.

Results for the IRF estimator based on an autoregression with order $h$
selected via AIC (i.e., $h=\hat{h}_{AIC}$) (Panel A) tell a qualitatively
similar story. Specifically, we find that the raw sieve generally still
performs well, with two exceptions, both of which occur for $d$, $\phi $ and
$k$ combinations for which the unadjusted estimator happens to be already
effectively unbiased. The pre-filtered method does better as the lag-length
increases, and best for high persistence ($d=0.4,\phi =0.9$). The RMSE of
the bias-adjusted statistics, as before, is either comparable to the
unadjusted, or somewhat improved; with the pre-filtered technique resulting
in a reduction of up to 17\% in the higher persistence case. Indeed, for
this high persistence setting, results (not reported) for the full set of $k$
values $1,\ldots ,99$ demonstrate a considerable reduction overall in bias
for the bias-adjusted estimator based on the pre-filtering, relative to the
bias-adjusted estimator based on the raw sieve.

We conclude this section by noting that, while results for $d=0$ were
produced, for reasons of space they have not been included in the tables. In
brief, the performance of the raw sieve for $d=0$ (based on both values of $%
h $) is similar to its performance for $d=0.2$; namely, it produces some
reduction in bias, over and above the unadjusted estimator, although in this
case at the cost of a small overall increase in the RMSE. The application of
pre-filtering has a generally negative impact on performance, as might be
expected, given that the pre-filtering introduces a completely unnecessary
layer of estimation uncertainty into the exercise. However, given the
well-documented \textit{upward} bias of semi-parametric estimates of $d$
when long memory is absent -- see, for example, \cite%
{agiakloglou:newbold:wohar:1993}, \cite{lieberman:2001b} and \cite%
{poskitt:martin:grose:2014} -- plus the downward bias in the persistence
measures that is documented in the current paper (and that continues to
obtain when $d=0$), conventional preliminary analysis is unlikely to lead a
researcher to conclude in favour of long memory when it is not present.
Hence, we would argue that it is unlikely that pre-filtering would ever be
invoked when $d=0$ and that the performance of the bias-adjusted estimates
based on the pre-filtered sieve in this setting has limited relevance for
empirical practice.

\subsubsection{Bias correction of the sample ACF}

As in the previous section, we begin by plotting selected distributional
results for the sample ACF, where the bias adjustment occurs via the raw
sieve algorithm. Panels (i) to (v) in each figure plot respectively: the
Monte Carlo distribution of the unadjusted statistic $\widehat{\rho }(k)$;
the Monte Carlo distribution of the bootstrap bias-adjusted statistic, $%
\widehat{\rho }^{(BA)}(k)$; the average bootstrap estimate of the
distribution of $\widehat{\rho }(k)$; and the Monte Carlo distribution of
the estimator adjusted using the (infeasible) asymptotic bias formula of
\cite{hosking:1996}, $\widehat{\rho }^{(ASY)}(k)$. The four plots are
indicated by the legend entries \textquotedblleft MC\textquotedblright ,
\textquotedblleft MC-BA\textquotedblright , \textquotedblleft
BS-av\textquotedblright\ and \textquotedblleft BA-asy\textquotedblright\
respectively. The vertical dotted line indicates the position of the true
value of $\rho (k)$ for each $k=1,3,6,9,12$. As previously noted, the
bootstrapping is performed in terms of the Fisher $z$ transform of the ACF
so as to restrict the bias-adjusted ACF to the $\left( -1,1\right) $
interval.

Panel (vi) plots, for lags $k=1,2,...,99$, the true ACF $\rho (k)$ (based on
the parameters of the true data generating process); the mean of the Monte
Carlo distribution of $\widehat{\rho }(k)$; and the mean of the Monte Carlo
distribution of $\widehat{\rho }^{(BA)}(k)$ (designated \textquotedblleft
True $\rho $\textquotedblright , \textquotedblleft $\overline{MC}$%
\textquotedblright , and \textquotedblleft $\overline{MC\text{-}BA}$%
\textquotedblright\ respectively). In Panel (i) we also plot the sampling
distribution of the feasible \citeauthor{lee:ko:2009} bias-adjusted
estimator (referred to hereafter as $\widehat{\rho }^{(LK)}(1)$, and
designated \textquotedblleft BA-LK\textquotedblright\ on the figure).

\begin{figure}[tbph]
\centering%
\includegraphics[trim=10mm 70mm 10mm
80mm,width=6in]{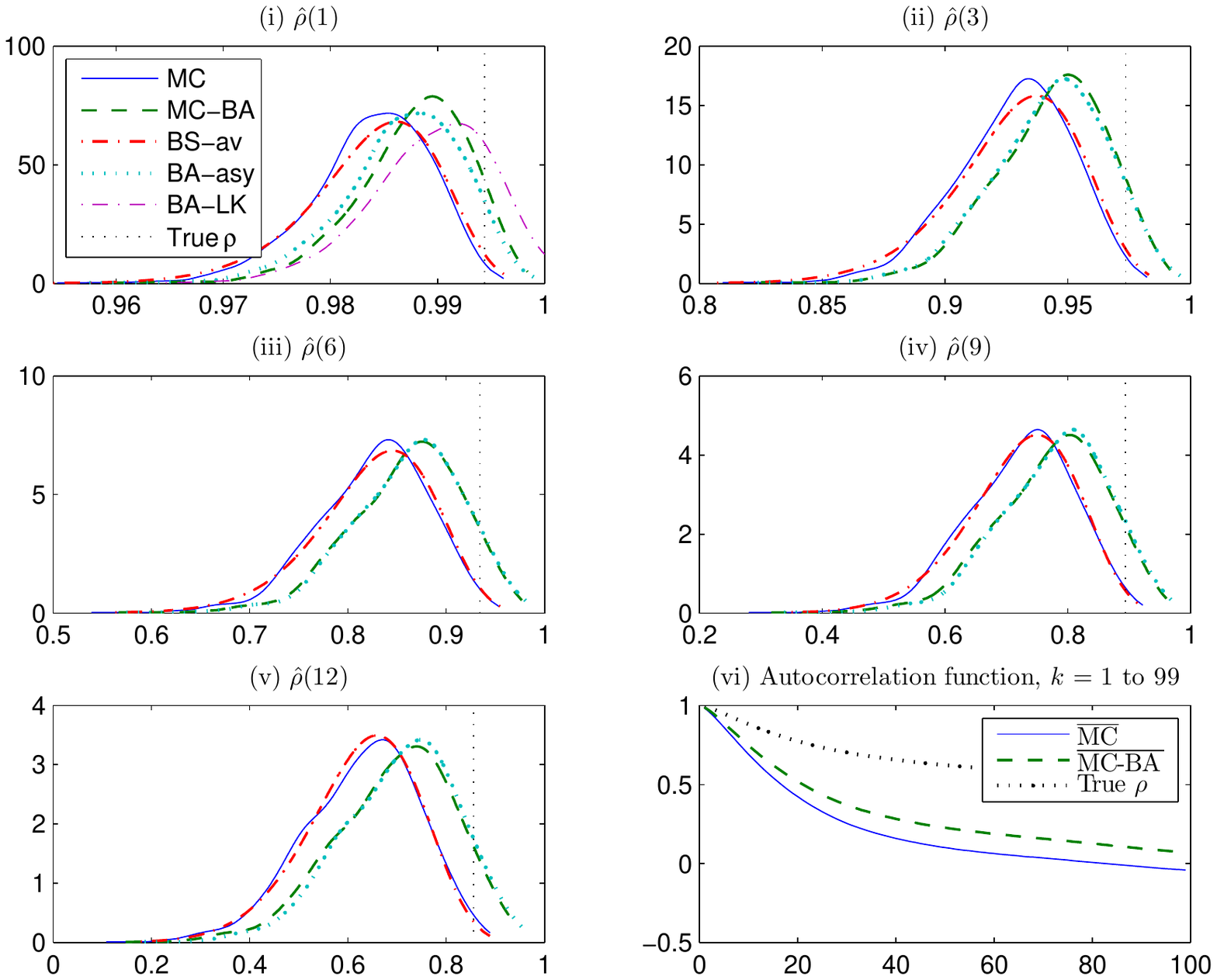}
\caption{Bias correction of the sample ACF using the \textit{raw sieve}
bootstrap.\protect\smallskip \newline
\textit{True process: }ARFIMA$(1,d,0)$; $T=500$; $d=0.4$; $\protect\phi =0.9$%
. \protect\smallskip \newline
\textit{Key for Panels (1) to (v): }\textquotedblleft MC\textquotedblright :
Monte Carlo distribution of the unadjusted\ statistic $\protect\widehat{%
\protect\rho }(k)$; \textquotedblleft MC-BA\textquotedblright : Monte Carlo
distribution of the bootstrap bias-adjusted statistic $\protect\widehat{%
\protect\rho }^{(BA)}(k)$; \textquotedblleft BS-av\textquotedblright : the
average bootstrap estimate of the distribution of $\protect\widehat{\protect%
\rho }(k)$; \textquotedblleft BA-asy\textquotedblright : the Monte Carlo
distribution of $\protect\widehat{\protect\rho }^{(ASY)}(k)$;
\textquotedblleft BA-LK\textquotedblright : the Monte Carlo distribution of $%
\protect\widehat{\protect\rho }^{(LK)}(1)$. \textit{Key for Panel (vi): }%
\textquotedblleft $\overline{MC}$\textquotedblright : mean of the Monte
Carlo distribution of $\protect\widehat{\protect\rho }(k)$;
\textquotedblleft $\overline{MC\text{-}BA}$\textquotedblright : mean of the
Monte Carlo distribution of $\protect\widehat{\protect\rho }^{(BA)}(k)$. The
true value of $\protect\rho (k)$ is indicated by the use of small dots in
all panels.}
\label{fig8}
\end{figure}

Figure \ref{fig8} displays distributional results as listed above for $T=500$%
, with $d=0.4$ and $\phi =0.9$. As was the case with IRF estimation we find
that the qualitative results for ACF estimation are robust to the method by
which $h$ is selected, with there being no clear superiority of one set of
results over the other. In this case we choose to present graphical results
for the more conventional choice of $h$, based on AIC, with results for $h=%
\left[ (\ln T)^{2}\right] $ reproduced in Table \ref{newtbl2} only. Once
again, corresponding results for $T=100$ can be found in the \href{http://users.monash.edu.au/~gmartin/Grose_Martin_Poskitt_on_line_appendix.pdf}%
{Supplementary Appendix} and are discussed explicitly here only when they differ
qualitatively from those for $T=500$.

\begin{figure}[tbph]
\centering%
\includegraphics[trim=10mm 70mm 10mm
80mm,width=6in]{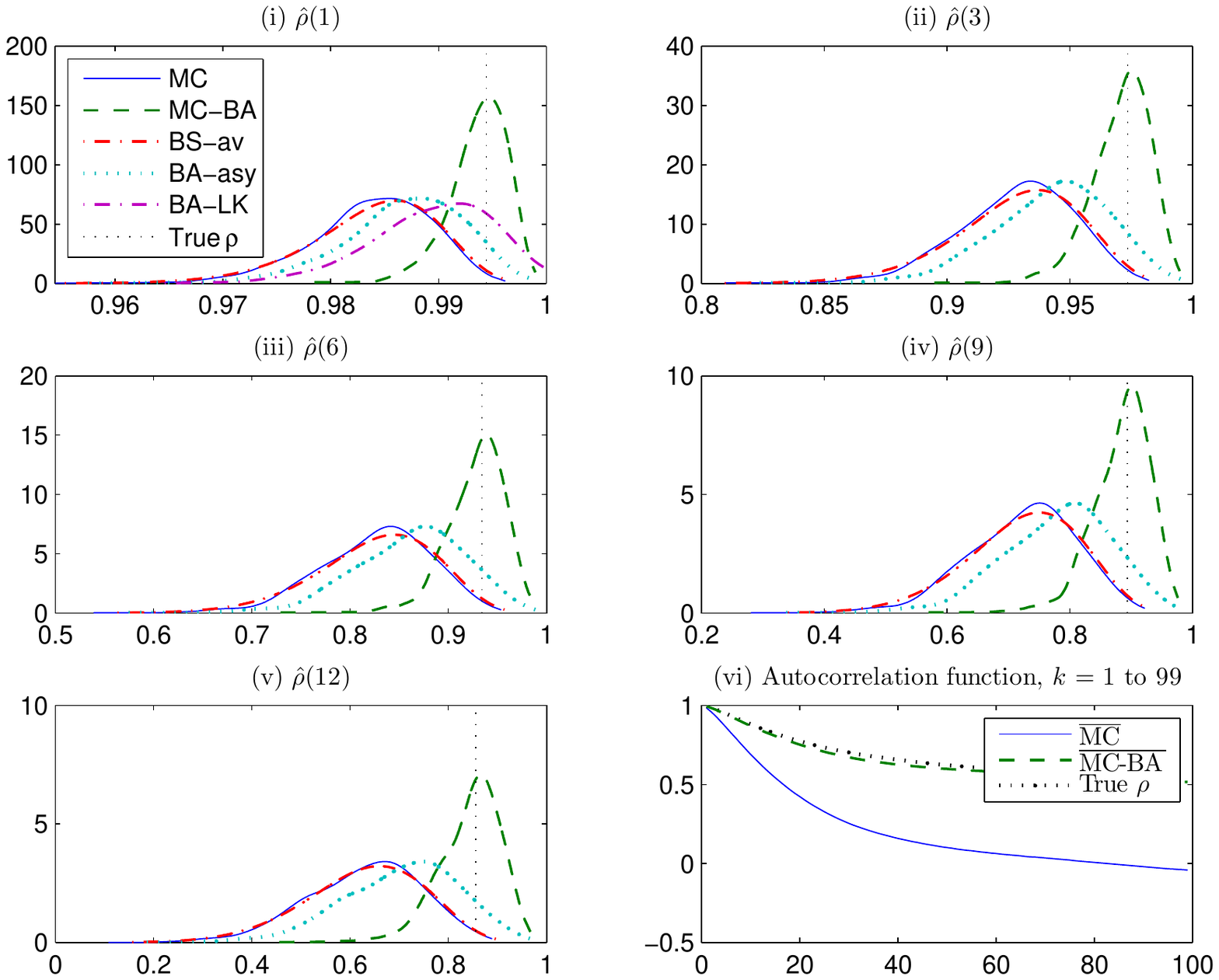}
\caption{Bias correction of the sample ACF using the \textit{pre-filtered
sieve} bootstrap, based on the \textit{true value of }$d$ \textit{as the
pre-filter}\protect\smallskip \newline
\textit{True process: }ARFIMA$(1,d,0)$; $T=500$; $d=0.4$; $\protect\phi =0.9$%
. \protect\smallskip \newline
\textit{Key for Panels (1) to (v): }\textquotedblleft MC\textquotedblright :
Monte Carlo distribution of the unadjusted\ statistic $\protect\widehat{%
\protect\rho }(k)$; \textquotedblleft MC-BA\textquotedblright : Monte Carlo
distribution of the bootstrap bias-adjusted statistic $\protect\widehat{%
\protect\rho }^{(BA)}(k)$; \textquotedblleft BS-av\textquotedblright : the
average bootstrap estimate of the distribution of $\protect\widehat{\protect%
\rho }(k)$; \textquotedblleft BA-asy\textquotedblright : the Monte Carlo
distribution of $\protect\widehat{\protect\rho }^{(ASY)}(k)$;
\textquotedblleft BA-LK\textquotedblright : the Monte Carlo distribution of $%
\protect\widehat{\protect\rho }^{(LK)}(1)$. \textit{Key for Panel (vi): }%
\textquotedblleft $\overline{MC}$\textquotedblright : mean of the Monte
Carlo distribution of $\protect\widehat{\protect\rho }(k)$;
\textquotedblleft $\overline{MC\text{-}BA}$\textquotedblright : mean of the
Monte Carlo distribution of $\protect\widehat{\protect\rho }^{(BA)}(k)$. The
true value of $\protect\rho (k)$ is indicated by the use of small dots in
all panels.}
\label{fig10}
\end{figure}

Largely mimicking the results pertaining to the estimation of the IRF, the
sieve-based technique reproduces quite accurately the `true'\ Monte Carlo
distribution of the statistic to be bias-adjusted. However, as Figure \ref%
{fig8} demonstrates, and as has been documented elsewhere (see, for example, %
\citealp{hosking:1996} and \citealp{poskitt:grose:martin:2013}), the
conventional autocorrelation coefficient $\widehat{\rho }(k)$ is \emph{very}
biased, and none of the techniques considered here manage to completely
eradicate that bias. The raw sieve bias-adjustment technique does,
nevertheless, succeed in producing a statistic $\widehat{\rho }^{(BA)}(k)$
that is notably \emph{less} biased than the unadjusted statistic.\ In fact,
for this sample size the sieve-based technique produces an estimate of $\rho
(k)$\ that is as accurate (for the recorded values of $k$) as the
analytically adjusted estimator, $\widehat{\rho }^{(ASY)}(k)$, based on the
the known data generating parameters! Making reference to the corresponding
figure for $T=100$ included in the appendix, for the smaller sample size the
sieve-based method is actually more accurate than the infeasible $\widehat{%
\rho }^{(ASY)}(k)$, with $\widehat{\rho }^{(BA)}(k)$ being both less biased
and having a much smaller RMSE than $\widehat{\rho }^{(ASY)}(k)$ on average.

Comparing $\widehat{\rho }^{(BA)}(1)$ with the bias-adjusted estimator $%
\widehat{\rho }^{(LK)}(1)$ based on the estimated \citeauthor{lee:ko:2009}
bias, we see that our `plug-in'\ estimate of the latter results in an
estimator with slightly less bias than that of $\widehat{\rho }^{(BA)}(1)$,
but at the cost of a slightly larger RMSE. Indeed, the results recorded for $%
T=100$ in the appendix demonstrate that for the smaller sample size the
dispersion of the sampling distribution of $\widehat{\rho }^{(LK)}(1)$ is
very large, rendering it an unreliable bias adjustment method in such a
setting.

In Figure \ref{fig10} we plot the corresponding results based on the
pre-filtered bootstrap technique, with the true value of $d$ used in the
pre-filtering. The results confirm, once again, the remarkable accuracy of
this approach, with the bias-adjusted estimator seen to have a mean value
(across Monte Carlo replications) that is almost visually indistinguishable
from the true $\rho (k)$\ for all values of $k$\ considered.

However, in contrast to the case for the IRF, rendering the\ pre-filtered
technique feasible via the substitution of the SPLW estimate for $d$ in the
pre-filtering algorithm does not produce a bias-adjusted estimator whose
performance mimics that of the estimator that exploits the true value of $d$%
. Instead, the procedure results in a severe over-correction of the Fisher-$%
z $\ transformed ACF which, when passed through the reverse transform (\ref%
{ifz}), results in coefficients that are biased towards one. The severity of
this over-correction naturally worsens as the degree of persistence
increases (i.e., as $d$\ and/or $\phi $\ increase), to the extent that, for
the highest persistence design considered, the bias-\textquotedblleft
corrected\textquotedblright\ estimates were all just less than one. For very
low values of $k$ this in fact leads to less bias, as we see from the
results recorded in Table \ref{newtbl2}. However, when considering the
results for the ACF as a whole, over the full spectrum of lags extending out
to $k=99$, the use of pre-filtering to bias correct is problematic, and
those results are not therefore documented graphically. Careful
investigation of the underlying outcomes indicates that the SPLW estimator
is itself biased upwards, and that the bias in the SPLW estimator of $d$
skews the reference value of $\rho (k)$ in such a way that its use as a
basis for calculating the bootstrap estimate of bias is severely
compromised. Thus, despite the accuracy of the estimate of the sampling
distribution of $\widehat{\rho }(k)$ as produced by the pre-filtered sieve
based on the true $d$, inaccuracy in the estimate of $d$ can produce a
reference value for use in the bias-correction that is itself an inaccurate
representation of the true but unknown value of $\rho (k)$ that underlies
the data generating process. Hence, the bootstrap-based measure of bias is
not an accurate estimate of the true unknown bias in $\widehat{\rho }(k)$.

\afterpage{
\begin{landscape}
\setlength{\oddsidemargin}{-0.4in}
\setlength{\textwidth}{6.8in}

\begin{table}[tbp] \centering\caption{
Bias and root mean squared error (RMSE) of estimators of selected
autocorrelation coefficients, for $T=500$.\\
Results for the unadjusted and both forms of bootstrap-based bias-adjusted
estimators are documented.}\label{newtbl2}

\begin{tabular}{ccccccccccccccccc}
&  &  &  &  &  &  &  &  &  &  &  &  &  &  &  &  \\[-2.5ex]
&  &  & \multicolumn{4}{c}{$\widehat{\rho }(k)$} &  & \multicolumn{4}{c}{$\widehat{\rho }^{(BA)}(k)$ (raw sieve)} &  & \multicolumn{4}{c}{$\widehat{\rho }^{(BA)}(k)$ (pre-filtered sieve)} \\
\cline{4-7}\cline{9-12}\cline{14-17}
&  &  &  &  &  &  &  &  &  &  &  &  &  &  &  &  \\[-1.5ex]
&  &  & $k=1$ & $k=6$ & $k=12$ & av. &  & $k=1$ & $k=6$ & $k=12$ & av. &  & $k=1$ & $k=6$ & $k=12$ & av. \\[2ex]
\multicolumn{17}{c}{Panel A: $T=500;$ $h$ based on AIC selection} \\[0.5ex] \hline
&  &  &  &  &  &  &  &  &  &  &  &  &  &  &  &  \\[-1ex]
$d$ & $\phi $ & \multicolumn{15}{c}{Bias} \\[1ex]
0.2 & 0.6 &  & -0.0166 & -0.0509 & -0.057 & -0.0437 &  & -0.0100 & -0.0327 &
-0.0394 & -0.0286 &  & 0.0993 & 0.3311 & 0.3964 & 0.2875 \\
& 0.9 &  & -0.0100 & -0.0667 & -0.1143 & -0.0636 &  & -0.0045 & -0.0315 &
-0.0555 & -0.0303 &  & 0.0322 & 0.2404 & 0.4409 & 0.2357 \\
0.4 & 0.6 &  & -0.0445 & -0.2191 & -0.2927 & -0.1912 &  & -0.0377 & -0.1889
& -0.2549 & -0.1653 &  & 0.0577 & 0.2913 & 0.3935 & 0.2548 \\
& 0.9 &  & -0.0106 & -0.1085 & -0.2283 & -0.1128 &  & -0.0066 & -0.0752 &
-0.1623 & -0.0790 &  & 0.0055 & 0.0652 & 0.1428 & 0.0690 \\
&  &  &  &  &  &  &  &  &  &  &  &  &  &  &  &  \\[-1ex]
&  & \multicolumn{15}{c}{RMSE} \\[1ex]
0.2 & 0.6 &  & 0.0336 & 0.1007 & 0.1065 & 0.0849 &  & 0.0311 & 0.0957 &
0.1011 & 0.0803 &  & 0.1249 & 0.4174 & 0.483 & 0.3578 \\
& 0.9 &  & 0.0146 & 0.0975 & 0.1661 & 0.0927 &  & 0.0114 & 0.0796 & 0.1381 &
0.0762 &  & 0.0322 & 0.2404 & 0.4409 & 0.2357 \\
0.4 & 0.6 &  & 0.0497 & 0.2429 & 0.3211 & 0.2112 &  & 0.0439 & 0.2186 &
0.2912 & 0.1904 &  & 0.0583 & 0.2940 & 0.3962 & 0.2570 \\
& 0.9 &  & 0.0120 & 0.1224 & 0.2564 & 0.1270 &  & 0.0085 & 0.0946 & 0.2028 &
0.0991 &  & 0.0055 & 0.0652 & 0.1428 & 0.0690 \\
&  &  &  &  &  &  &  &  &  &  &  &  &  &  &  &  \\
\multicolumn{17}{c}{Panel B: $T=500;$ $h=(\ln T)^{2}$} \\[0.5ex] \hline
&  &  &  &  &  &  &  &  &  &  &  &  &  &  &  &  \\[-1ex]
$d$ & $\phi $ & \multicolumn{15}{c}{Bias} \\[1ex]
0.2 & 0.6 &  & -0.0166 & -0.0509 & -0.0570 & -0.0437 &  & -0.0085 & -0.0271
& -0.0323 & -0.0237 &  & 0.0741 & 0.2560 & 0.3096 & 0.2221 \\
& 0.9 &  & -0.0100 & -0.0667 & -0.1143 & -0.0636 &  & -0.0047 & -0.0335 &
-0.0603 & -0.0326 &  & 0.0321 & 0.2397 & 0.4396 & 0.2350 \\
0.4 & 0.6 &  & -0.0445 & -0.2191 & -0.2927 & -0.1912 &  & -0.0354 & -0.1760
& -0.2370 & -0.1540 &  & 0.0570 & 0.2882 & 0.3898 & 0.2522 \\
& 0.9 &  & -0.0106 & -0.1085 & -0.2283 & -0.1128 &  & -0.0067 & -0.0767 &
-0.1664 & -0.0808 &  & 0.0055 & 0.0650 & 0.1423 & 0.0688 \\
&  &  &  &  &  &  &  &  &  &  &  &  &  &  &  &  \\[-1ex]
&  & \multicolumn{15}{c}{RMSE} \\[1ex]
0.2 & 0.6 &  & 0.0336 & 0.1007 & 0.1065 & 0.0849 &  & 0.0313 & 0.0978 &
0.1061 & 0.0827 &  & 0.1106 & 0.3752 & 0.4339 & 0.3209 \\
& 0.9 &  & 0.0146 & 0.0975 & 0.1661 & 0.0927 &  & 0.0116 & 0.0815 & 0.1446 &
0.0788 &  & 0.0321 & 0.2397 & 0.4396 & 0.2350 \\
0.4 & 0.6 &  & 0.0497 & 0.2429 & 0.3211 & 0.2112 &  & 0.0425 & 0.2110 &
0.2818 & 0.1840 &  & 0.0580 & 0.2925 & 0.3939 & 0.2556 \\
& 0.9 &  & 0.0120 & 0.1224 & 0.2564 & 0.1270 &  & 0.0086 & 0.0964 & 0.2083 &
0.1014 &  & 0.0055 & 0.0650 & 0.1423 & 0.0688 \\
&  &  &  &  &  &  &  &  &  &  &  &  &  &  &  &
\end{tabular}
\end{table}
\end{landscape}}

To understand this point it is worth remembering the situation that obtains
for bias-adjustment in a parametric bootstrap setting. In that case, an
unknown parameter $\theta $ that characterizes the data generating process
is estimated as $\widehat{\theta }$. Repeated bootstrap samples are then
generated from the estimated model (based on $\widehat{\theta })$, producing
repeated bootstrap values, $\widehat{\theta }_{b},b=1,2,...,B,$ and the bias
of $\widehat{\theta }$, defined as $E(\widehat{\theta }-\theta )$, is
estimated by $\frac{1}{B}\textstyle\sum_{b=1}^{B}\widehat{\theta }_{b}-%
\widehat{\theta }.$ The key here is that $\widehat{\theta }$ plays exactly
the same role in generating the bootstrap samples as does $\theta $ in
generating the empirical sample. In the case of bootstrapping the IRF or ACF
using the (pre-filtered) sieve, however, the true data generating process is
(by the very nature of the exercise) \textit{not} estimated but, rather,
approximated via the combination of an estimate of $d$ and the fitted
autoregression. The requisite parameter reference values to use in the
bootstrap bias calculations therefore need to be backed out from the
approximating model.

Now, whereas inaccuracies in the estimate $\hat{d}$ appear to be compensated
for by changes in the autoregressive estimates $\hat{\phi}_{h}(1),\ldots ,%
\hat{\phi}_{h}(h)$ of the $AR(h)$ approximation fitted to the filtered data $%
(1-z)^{\hat{d}}y(t)$, in such a way that the reference IRF $\tilde{\psi}(k)$
implicit in the $ARFMA(h,\hat{d},0)$ approximating model provides a clear
reflection of the true IRF coefficients; the same is not true of the ACF. A
small amount of inaccuracy in the estimate of $d$ produces an implied
reference value $\tilde{\rho}(k)$ that is sufficiently different from what
would be produced by using the true (unknown) value of $d$ to ultimately
produce an inaccurate estimate of the true bias of $\widehat{\rho }(k)$. The
reason for this difference in sensitivity presumably lies in the fact that
for any given values of $\hat{d}$ and $\hat{\phi}_{h}(1),\ldots ,\hat{\phi}%
_{h}(h)$ the reference values for the two different statistics are related
via the expression $\tilde{\rho}(k)=\sum_{s\geq k}\tilde{\psi}(s)\tilde{\psi}%
(k-s)/\sum_{s\geq 0}\tilde{\psi}^{2}(s)$. This suggests that small
perturbations in the $\tilde{\psi}(k)$, that are immaterial for the
pre-filtered-based bias correction of $\widehat{\psi }(k)$, multiply and
accumulate so as to result in a change in the value of $\tilde{\rho}(k)$
that is sufficiently large to distort the corresponding bias correction of $%
\widehat{\rho }(k)$. The implication is that use of the pre-filtered sieve
to bias correct the ACF requires a greater degree of precision in the
preliminary estimate $\hat{d}$ in order to achieve the high level of
accuracy seen when employing the method to bias correct the IRF. Whilst it
is beyond the scope of this paper to investigate this point further, we note
that in related work \citep{poskitt:martin:grose:2014} the authors are
investigating the use of sieve-based techniques to bias adjust $d$ itself.
It could be hoped that such a procedure may produce estimates of $d$ that
are accurate enough to alleviate the sensitivity problem observed here in
the bias adjustment of $\widehat{\rho }(k)$. We leave that investigation for
a later date.

As a final point, results produced (but not included here) for $d=0 $\ show
that, in common with the IRF results, the use of the raw sieve to bias
adjust the ACF in this setting continues to yield a reduction in bias. In
contrast with the IRF results, however, this reduction in bias is also
sometimes sufficient to produce a reduction in RMSE. Once again, redundant
pre-filtering does not yield improvements overall.

\section{Discussion}

This paper has demonstrated the benefits of using bootstrap techniques to
reduce the bias of the primary persistence measures -- the autocorrelation
and impulse response functions -- in long memory settings. Given the
difficulty of accurately specifying the short memory dynamics in long memory
ARFIMA models, a semi-parametric approach to the bootstrap has been adopted,
with pre-filtering based on a preliminary semi-parametric estimate of the
long memory parameter also advocated. The results provide quite clear
guidance for the researcher wishing to draw conclusions about persistence in
this setting. The fact that the raw sieve yields bias improvements at
little, if any, cost in RMSE for both persistence measures in virtually all
settings, including those in which long memory is actually absent, leads us
to recommend that the raw sieve should be used as the default method for
bias adjustment. In the case of the impulse response function, if the
preliminary evidence in favour of long memory is reasonably strong, the
pre-filtered sieve should definitely be invoked, knowing that the extent of
the extra bias adjustment so produced can be substantial. Comparison of the
pre-filtering method with an alternative approach based on a modification of
the \cite{kilian:1998} technique for bias adjusting the impulse response
function serves to confirm this conclusion, with the pre-filtered sieve
yielding results that are either comparable or better, at no extra
computational burden. In the case of the autocorrelation function, the
results indicate that a very accurate estimate of the pre-filter is required
if the pre-filtering technique is to be reliable as a method of bias
adjustment for all lag values, and under any true settings.

Finally, we reiterate that the scope of our paper has been restricted to
using the bootstrap to bias-adjust persistence measures, and measuring the
accuracy of the estimators so produced via conventional means. As noted in
the Introduction, some attention in the literature has been given to the use
of the bootstrap to improve the accuracy of confidence intervals for impulse
response functions in particular, in time series settings that do encompass
long memory processes. Further work in this direction is the subject of
ongoing research.

\phantomsection\addcontentsline{toc}{section}{Supplementary Material}
\subsection*{Supplementary Material}

The additional Tables and Figures referenced in Section \ref{simres} can be accessed
at %
\url{http://users.monash.edu.au/~gmartin/Grose_Martin_Poskitt_on_line_appendix.pdf}

\subsection*{Acknowledgements}

The authors would like to thank two anonymous referees for very constructive
and helpful comments on an earlier draft of the paper. This research has
been supported by Australian Research Council (ARC) Discovery Grant
DP120102344 and ARC Future Fellowship FT0991045.

\phantomsection\addcontentsline{toc}{section}{References}
\bibliographystyle{ims}
\bibliography{tsa}

\end{document}